\documentclass[aps,prx,twocolumn,superscriptaddress,floatfix,letter]{revtex4-2}
\usepackage{epsfig,amsmath,amssymb,color,amsfonts,physics,microtype}
\usepackage[bookmarks=true,colorlinks,citecolor=red]{hyperref}
\usepackage{dsfont}
\usepackage{wrapfig}
\usepackage{tikz}
\usepackage{physics}
\usepackage{mathrsfs}
\usepackage[export]{adjustbox}
\usepackage{soul}
\usepackage{comment}
\usepackage{subfigure}
\usepackage{amsthm}
\usepackage{mathrsfs}
\usepackage{bbold}
\usepackage{orcidlink}

\newcommand{\Id}{\hat{\mathbb{1}}}
\newcommand{\PauliX}{{\hat{\text{X}}}}
\newcommand{\PauliY}{{\hat{\text{Y}}}}
\newcommand{\PauliZ}{{\hat{\text{Z}}}}

\begin{document}

\title{Magic phase transitions in monitored gaussian fermions}

\author{Emanuele Tirrito~\orcidlink{0000-0001-7067-1203}}
\email{etirrito@ictp.it}
\affiliation{The Abdus Salam International Centre for Theoretical Physics (ICTP), Strada Costiera 11, 34151 Trieste, Italy}
\affiliation{Dipartimento di Fisica ``E. Pancini", Universit\`a di Napoli ``Federico II'', Monte S. Angelo, 80126 Napoli, Italy}

\author{Luca Lumia}
\affiliation{International School for Advanced Studies (SISSA), via Bonomea 265, 34136 Trieste, Italy}

\author{Alessio Paviglianiti}
\affiliation{International School for Advanced Studies (SISSA), via Bonomea 265, 34136 Trieste, Italy}

\author{Guglielmo Lami~\orcidlink{0000-0002-1778-7263}}
\affiliation{Laboratoire de Physique Th\'eorique et Mod\'elisation, CNRS UMR 8089, CY Cergy Paris Universit\'e, 95302 Cergy-Pontoise Cedex, France}

\author{Alessandro Silva}
\affiliation{International School for Advanced Studies (SISSA), via Bonomea 265, 34136 Trieste, Italy}

\author{Xhek Turkeshi~\orcidlink{0000-0003-1093-3771}}
\affiliation{Institute f\"{u}r Theoretische Physik, Universit\"{a}t zu K\"{o}ln, Z\"{u}lpicher Straße 77, D-50937 K\"{o}ln, Germany}

\author{Mario Collura~\orcidlink{0000-0003-2615-8140}}
\email{mcollura@sissa.it}
\affiliation{International School for Advanced Studies (SISSA), via Bonomea 265, 34136 Trieste, Italy}
\affiliation{INFN Sezione di Trieste, 34136 Trieste, Italy}

\begin{abstract}
Monitored quantum systems, where unitary dynamics compete with continuous measurements, exhibit dynamical transitions as the measurement rate is varied. 
These reflect abrupt changes in the structure of the evolving wavefunction, captured by complementary complexity diagnostics that include and go beyond entanglement aspects. 
Here, we investigate how monitoring affects magic state resources—the nonstabilizerness—of Gaussian fermionic systems. Using scalable Majorana sampling techniques, we track the evolution of stabilizer Rényi entropies in large systems under projective measurements. While the leading extensive (volume-law) scaling of magic remains robust across all measurement rates, we uncover a sharp transition in the subleading logarithmic corrections. This measurement-induced complexity transition, invisible to standard entanglement probes, highlights the power of magic-based diagnostics in revealing hidden features of monitored many-body dynamics. 
\end{abstract}

\maketitle

\section{Introduction}
The complexity of a quantum state can be characterized in many ways, the most common being related to bipartite entanglement. While entanglement is a fundamental resource for quantum computation~\cite{RevModPhys.80.517, Chitambar_2019,Gour_2024}, it is however not the sole measure of quantum complexity. Different information is obtained for example by the non-stabilizerness or ``magic'',  which quantifies how much a quantum state deviates from being a stabilizer state~\cite{Gottesman_1997, Gottesman_1998_1, Gottesman_1998_2}, a state generated by quantum circuits consisting solely of Clifford operations which can be efficiently simulated on a classical computer~\cite{Valiant_2001, Dehaene_2003,Aaronson_2004}.

Since Clifford operations do generate entanglement, it is indeed evident that the complexity associated with non-stabilizerness refers to other aspects of quantum states~\cite{Winter_2022}. Non-stabilizerness, often considered in the context of quantum error correction and fault-tolerant quantum computing~\cite{Howard_2017}, provides new insights into the deeper structure of quantum systems and their computational capabilities. This perspective has sparked considerable interest within the scientific community,  as reflected by the increasing volume of recent research devoted to many-body quantum magic~\cite{Oliviero_2022,Odavic_2023, Tarabunga_2023_1, Turkeshi_2023_2, Tarabunga_2024,Tarabunga_2024_3,hoshino2025stabilizerrenyientropyconformal,Frau_2024_1,Frau_2024_2,Turkeshi_2024_2,zhang2024quantummagicdynamicsrandom,hou2025stabilizerentanglementmagichighway,santra2025complexitytransitionschaoticquantum, tirrito2024anticoncentrationmagicspreadingergodic,szombathy2025independentstabilizerrenyientropy,szombathy2025spectralpropertiesversusmagic,sticlet2025nonstabilizernessopenxxzspin,Leone2021quantumchaosis,tirrito2025universalspreadingnonstabilizernessquantum,turkeshi2025_c,magni2025anticoncentrationcliffordcircuitsbeyond,magni2025quantumcomplexitychaosmanyqudit,haug2024probingquantumcomplexityuniversal, Catalano_2024,Passarelli_2024_1,Passarelli_2024_2,Passarelli_2025,Jasser_2025,Odavic_2025,Bera_2025,Sinibaldi_2025}.

A recent line of research in quantum information and condensed matter physics focuses on monitored quantum systems. These are systems that, while evolving, are subject to random measurements that continuously alter their state. Such dynamics has been studied both in the context of discrete quantum circuits and for continuously monitored Hamiltonian systems, see~\cite{li2025measurementinducedentanglementphasetransition,Fisher_2023,Potter_2022} for a review. When the system is subject to continuous measurements, the mean evolution of the quantum state is typically described by a Lindblad equation, which emerges as an average over different quantum trajectories.  
Studying the properties of individual quantum trajectories provides a different perspective on the so-called ``measurement-induced'' dynamics and can lead to surprising and non-trivial effects on the system’s entanglement properties. Specifically, the interplay between the entangling unitary evolution and the effects of random measurements (which may entangle or disentangle depending on the specifics~\cite{Paviglianiti_2024}) can give rise to different regimes of entanglement scaling with system size, giving rise to the phenomenon of entanglement transitions~\cite{Cao_2019, Skinner_2019,Li2019,Alberton_2021,Turkeshi_2021,Turkeshi_2022,Coppola_2022,Muller_2022,Merritt_2022,ippoliti2020mipt,Tirrito_2023,Biella_2021,choi2020mipt,gullans2020mipt,gullans2020scalable,xing2024,jian2020mipt,Turkeshi_2020,Coppola_2022,Fava_2023,bucchold2022mipt,buchhold2024mipt,Li2025,Lami_2024_1}.

The study of transitions in monitored quantum systems has recently been extended in the broader framework of quantum complexity~\cite{Sierant_2022}. In Clifford circuits subject to a competition between repeated measurements and magic injection via $T$-gates, sharp transitions can emerge, not only in entanglement properties but also in different measures of non-stabilizerness~\cite{Fux_2024, Paviglianiti_2024, Liu_2024,Tarabunga_2024_4}. This can be expected by observing that, for example, local projective measurements of Pauli strings collapse the state into one stabilized by the measured operator. Therefore, the injection of T-gates and local measurements are two mechanisms driving the system towards opposite limits. Early investigations largely focused on quantum circuits with random measurements, shedding light on entanglement transitions but providing only a glimpse of the deeper interplay between monitoring and quantum complexity. The evolution of non-stabilizerness under Hamiltonian dynamics with continuous monitoring has only recently begun to be addressed, with existing studies limited to small system sizes~\cite{Bejan_2024_1, Russomanno_2025}. Pushing these analyses toward larger systems and the thermodynamic limit is therefore crucial for uncovering the true nature of measurement-induced complexity transitions.

In this work, we study the behavior of non-stabilizerness in monitored Gaussian quantum systems, a class that has been widely studied for its analytical tractability and its relevance in both quantum information theory and condensed matter physics. Gaussian states, fully specified by their first and second moments, naturally lend themselves to such analyses. The dynamics of Gaussian systems can either exhibit a $U(1)$ symmetry, as in models of hopping fermions, or only a $\mathbb{Z}_2$ parity symmetry, as seen in more general Gaussian circuits and Hamiltonian models like the Ising chain. Their structural simplicity and broad relevance make Gaussian systems an ideal setting to explore how measurement-induced dynamics impact quantum complexity. In particular, this framework offers new insights on non-stabilizerness, complementing the well-studied domain of entanglement.

The non-stabilizerness of these states is quantified through the Stabilizer Rényi Entropies (SREs), which provide a measure of how much a quantum state deviates from the stabilizer subspace~\cite{Leone_2022,Leone_2024}. These quantities are challenging to compute directly for large systems~\cite{Chen_2022, Haug_2023, Lami_2023_2, Lami_2024, Tarabunga_2024_2,Sinibaldi_2025,Tarabunga_2025}, but recent advances in Majorana sampling techniques provide a powerful tool for handling fermionic Gaussian states, allowing the computation of SREs even for systems with hundreds of lattice sites~\cite{Collura_2025}. This efficiency enables us to explore how the SREs scale with system size and how the measurement rate impacts the transition between different dynamical regimes.

In this article, by focusing on the scaling of the SREs with the system size, we show that
the transitions in the quantum system's complexity are effectively encoded in the presence of logarithmic corrections, which disappear beyond a critical measurement rate. This scaling provides a precise tool to study such transitions, as it directly reflects changes in the underlying quantum state’s “magic” and entanglement properties.
The paper is organized as follows. In Sec.~\ref{summary} we summarize the main findings of this work. In Sec.~\ref{sec:gaussian}, we review the formalism of fermionic Gaussian states. Section~\ref{sec:SREs_sampling} introduces SREs and their evaluation via Majorana sampling. In Sec.~\ref{sec:monitored_dynamics}, we present our results on monitored dynamics, focusing separately on non-interacting hopping fermions (Sec.~\ref{sec:hopping}) and the quantum Ising chain (Sec.~\ref{sec:ising}). We conclude with a discussion and outlook in Sec.~\ref{sec:conclusion}. Additional insights into the emergence of stationary non-stabilizerness, interpreted from the perspective of relaxation towards a generalized Gibbs ensemble, are provided in Appendix~\ref{sec:appendix_GGE}.

\section{Summary of Results}~\label{summary}

In this work, we investigate the emergence of complexity transitions in monitored Gaussian fermionic systems by analyzing the behavior of the SREs. Our analysis is based on large-scale numerical simulations that employ scalable Majorana sampling techniques to evaluate the SREs~\cite{Collura_2025}. These methods allow us to efficiently explore large quantum systems subjected to projective measurements.
Below, we summarize our main findings.
\begin{itemize}
    \item \textbf{Extensive Non-Stabilizerness}. In the absence of measurements, the SREs saturate to values that scale extensively with system size. Notably, in certain dynamical sectors—such as half-filling—the leading value agrees with the Haar average, indicating near-maximal quantum complexity in these regimes.

    \item \textbf{Sub-leading Logarithmic Corrections}. We identify logarithmic corrections to the extensive scaling of SREs, depending crucially on the initial state and the Rényi index. These corrections serve as a sensitive diagnostic tool for complexity transitions in monitored circuits.
    
    \item \textbf{Measurement-induced Transition}. When local projective measurements are introduced, a striking transition occurs in the logarithmic corrections to non-stabilizerness. As we show, above a critical measurement rate, these subleading logarithmic terms vanish abruptly. This signals a subtle yet sharp transition in quantum complexity, distinct from those typically revealed by entanglement measures.

    \item \textbf{Model-specific Behavior}. By analyzing two distinct Hamiltonian systems—the hopping fermions with particle-number conservation and the quantum Ising chain without such symmetry—we demonstrate that the identified transitions in non-stabilizerness are robust across different physical setups. 
    However, we observe subtle differences stemming from the roles of symmetries and conserved quantities.  
    
    \item \textbf{Finite-size and Time Evolution}. We thoroughly characterize the finite-size behavior of SREs and their dynamical approach to stationarity. This analysis reveals that, in the thermodynamic limit, the relaxation towards stationary non-stabilizerness typically exhibits algebraic decay, but finite-size effects can introduce exponential transient behaviors.
\end{itemize}

In conclusion, our results uncover novel stabilizerness phase transitions in monitored Gaussian fermionic systems, extending our understanding of measurement-induced dynamics and quantum complexity beyond conventional entanglement-based frameworks.

\section{Fermionic Gaussian States}\label{sec:gaussian}
Let us start considering a system of $L$ qubits, with Hilbert space dimension $D = 2^L$. Operators are expanded in the orthonormal basis of Pauli strings $\hat{\pmb{P}} \in \{ \Id, \PauliX, \PauliY, \PauliZ \}^{\otimes L}$, satisfying
$
\rm{Tr}( \hat{\pmb{P}}' \hat{\pmb{P}}) = D \, \delta_{\pmb{P}' \pmb{P}}.
$ 
Through the Jordan-Wigner transformation, we introduce $2L$ Majorana operators
$\hat{\gamma}_{2i-1} = \PauliZ_1 \cdots \PauliZ_{i-1} \PauliX_i$, $\hat{\gamma}_{2i} = \PauliZ_1 \cdots \PauliZ_{i-1} \PauliY_i$, which are Hermitian and satisfy the canonical anti-commutation relations $\{ \hat{\gamma}_\mu, \hat{\gamma}_\nu \} = 2 \delta_{\mu\nu}$.
The Majorana operators can alternatively be expressed using the fermionic creation and annihilation operators $\hat{c}_i$ and $\hat{c}_i^{\dag}$, as $\hat{\gamma}_{2i} = (\hat{c}_i^{\dag} + \hat{c}_i)$ and $\hat{\gamma}_{2i-1} = i(\hat{c}_i^{\dag} - \hat{c}_i)$.
Majorana monomials $\hat{\gamma}^{\pmb{x}} = (\hat{\gamma}_{1})^{x_1} (\hat{\gamma}_{2})^{x_2} ... (\hat{\gamma}_{2L})^{x_{2L}}$, labeled by binary strings $\pmb{x} \in \{0,1\}^{2L}$, form an orthonormal basis for operators as well, with $\rm{Tr}[ (\hat{\gamma}^{\pmb{x}'})^{\dagger} \hat{\gamma}^{\pmb{x}} ] = D \, \delta_{\pmb{x}' \pmb{x}}$.

A Gaussian unitary $\hat{U}$ is defined as a unitary operator that rotates Majorana operators linearly, according to
$
\hat{U}^{\dagger} \hat{\gamma}_\mu \hat{U} = \sum_{\nu=1}^{2L} O_{\mu\nu} \, \hat{\gamma}_\nu,
$
where $O \in SO(2L)$ is a real orthogonal matrix. 
Generic Gaussian density matrices take the form
\begin{equation}
\hat{\rho} = \hat{U}^{\dagger} \bigotimes_{i=1}^L \left( \frac{\Id + \lambda_i \PauliZ_i}{2} \right) \hat{U},
\end{equation}
where $\lambda_i \in [-1,1]$, and $\hat{U}$ is a Gaussian unitary. Pure Gaussian states correspond to the special case $\lambda_i = 1$ for all $i$.
Gaussian states are fully characterized by their covariance matrix, defined as
\begin{equation}
\mathbb{\Gamma}_{\mu\nu}(\hat\rho) = -\frac{i}{2} \, \text{Tr} \left( [\hat{\gamma}_\mu, \hat{\gamma}_\nu] \hat{\rho} \right),
\end{equation}
which evolves according to $\mathbb{\Gamma}(\hat{U}^{\dagger} \hat{\rho} \hat{U}) = O \mathbb{\Gamma}(\hat{\rho}) O^T$ under a Gaussian unitary.

The covariance matrix can be conveniently written in terms of the standard fermionic correlation matrix $\mathbb{C}_{\mu \nu} (\hat{\rho}) = \rm{Tr}( \mathbb{\hat c}_{\mu}^{\dag} \mathbb{\hat c}_{\nu}  \hat\rho)$,
where $\hat{\mathbb{c}}$ is defined as $\hat{\mathbb{c}}_{2i-1} = \hat{c}_i$ and $\hat{\mathbb{c}}_{2i} = \hat{c}_i^{\dag}$. Specifically, the covariance matrix is given by $\mathbb{\Gamma} = -i \left( 2 \Omega^* \mathbb{C} \Omega^T - \mathbb{1} \right)$, where $\Omega = \bigoplus_{i=1}^{L} \Omega^{(i)}$ is the unitary matrix that transforms the $\hat{\mathbb{c}}$ operators into the $\hat{\gamma}$ operators~\cite{Surace_2022}.
Let us briefly mention as well that particle-number-preserving Gaussian unitaries are defined as those unitaries for which the orthogonal matrix $O$ satisfies $[O, \mathbb{\Gamma}_0] = 0$, where $\mathbb{\Gamma}_0$ is the covariance matrix of the vacuum state
$\mathbb{\Gamma}_0 = \bigoplus_{i=1}^L 
\begin{bmatrix} 
0 & 1 \\ 
-1 & 0 
\end{bmatrix}
$.
Expectation values of Majorana monomials over Gaussian states are computed via Wick's theorem, and read
\begin{equation}
\text{Tr}\left( \hat{\rho} \, \hat{\gamma}^{\pmb{x}} \right) = i^{|\pmb{x}|/2} \, \text{Pf}\left( \mathbb{\Gamma}|_{\pmb{x}} \right),
\end{equation}
where $|\pmb{x}|$ is the Hamming weight of $\pmb{x}$ and $\mathbb{\Gamma}|_{\pmb{x}}$ denotes the submatrix of $\mathbb{\Gamma}$ restricted to the support of $\pmb{x}$, i.e., where all row and columns corresponding to the zero entries of $\pmb{x}$ are discarded. 

\section{Stabilizer Rényi Entropies and Majorana Sampling}\label{sec:SREs_sampling}
For $\alpha > 0$, the $\alpha$-SRE is defined as~\cite{Leone_2022}
\begin{equation}
    M_{\alpha}(\hat{\rho}) = \frac{1}{1-\alpha} \log \sum_{\pmb{P}} \pi_{\rho}^{\alpha}(\pmb{P}) - \log D \, ,
\end{equation}
which, up to an additive constant, coincides with the $\alpha$-Rényi entropy of the distribution
\begin{equation}
\pi_{\rho}(\pmb{P}) \equiv \frac{1}{D} \frac{\rm{Tr} [\hat{\pmb{P}} \hat{\rho}]^2}{\rm{Tr}[\hat{\rho}^2]}.
\end{equation}
As $\alpha \to 1$, the SRE reduces to the Shannon entropy
$M_1(\hat{\rho}) = - \sum_{\pmb{P}} \pi_{\rho}(\pmb{P}) \log \pi_{\rho}(\pmb{P}) - \log D$. SREs are useful in characterizing the distribution $\pi_{\rho}$ and have been shown to be effective measures of nonstabilizerness. Nonstabilizerness refers to the resource required for quantum states that cannot be reached by Clifford circuits, which are considered ``free'' operations, while non-Clifford gates are the resources. The Clifford group includes unitary transformations $\hat{U}_c$ that map Pauli strings to Pauli strings under conjugation. Pure stabilizer states, denoted as STAB, are those derived from the reference state $\ket{0 \dots 0}$ by Clifford transformations, carrying no resource. SREs, particularly for $\alpha \geq 2$, are considered effective measures of nonstabilizerness, with monotonicity rigorously established~\cite{Leone_2024, Haug_2023_2}. They also provide useful bounds for other measures of magic~\cite{Leone_2024}.

Let us now see how focusing on Gaussian states simplifies the computation of the SRE. Following Ref.~\cite{Collura_2025}, we start by observing that Majorana monomials and Pauli strings are in one-to-one correspondence. Consequently, the probability distribution
\begin{equation}
\pi_{\rho}(\pmb{x}) = \frac{1}{D} \frac{|\rm{Tr}[\hat{\rho}\,\hat\gamma^{\pmb{x}}]|^2}{\rm{Tr}[\hat{\rho}^2]}
\end{equation}
is equivalent to the distribution $\pi_{\rho}(\pmb{P})$ defined over Pauli strings. By applying Wick's theorem and using the identity that the square of a Pfaffian equals the determinant, we can further express it as
\begin{equation}\label{eq:dpp_1}
    \pi_{\rho}(\pmb{x}) = \frac{\det[\mathbb{\Gamma}|_{\pmb{x}}]}{\det[\mathbb{1} + \mathbb{\Gamma}]} \, ,
\end{equation}
where $\mathbb{\Gamma}$ denotes the covariance matrix of the fermionic Gaussian state, and $\mathbb{\Gamma}|_{\pmb{x}}$ represents its submatrix restricted to the Majorana modes selected by $\pmb{x}$. The denominator corresponds to the well-known expression for the purity of a fermionic Gaussian state~\cite{Surace_2022}, up to a factor containing the Hilbert space dimension.

It turns out that, in order to compute the SREs of a fermionic Gaussian state, one can first sample the distribution $\pi_{\rho}(\pmb{x})$ with $\mathcal{S}$ $2L$-bit strings $\pmb{x}$ and then evaluate the statistical average
\begin{eqnarray}
\mathbb{E}_{\pmb{x} \sim \pi_{\rho}(\pmb{x})} [\pi_{\rho}(\pmb{x})^{\alpha-1}]
\approx \frac{1}{\mathcal{S}}\sum_{\pmb{x} \sim \pi_{\rho}(\pmb{x})}^{\mathcal{S}}
\pi_{\rho}(\pmb{x})^{\alpha-1},
\end{eqnarray}
where $\pmb{x} \sim \pi_{\rho}(\pmb{x})$ means that $\pmb{x}$ is extracted with probability $\pi_{\rho}(\pmb{x})$; 
from that we then have $M_\alpha(\hat{\rho})\approx \frac{1}{1-\alpha}\log \mathbb{E}_{\pmb{x} \sim \pi_{\rho}(\pmb{x})} [\pi_{\rho}(\pmb{x})^{\alpha-1}] - \log D$.
The sampling is typically problematic because the full set of Majorana strings has exponential size $D^2=2^{2L}$. This difficulty can be overcome by decomposing the full probability into a product of conditional probabilities $\pi_{\rho}(\pmb{x})
= \pi_{\rho}(x_1)
\pi_{\rho}(x_2|x_1)
\cdots
\pi_{\rho}(x_{2L}|x_1\cdots x_{2L-1})$
and sampling the distribution iteratively, one bit at a time. This is effectively possible because the conditional probabilities $\pi_{\rho}(x_{\mu}|x_1\cdots x_{\mu-1})$ can be efficiently computed. To this end, the marginal probabilities $\pi_{\rho}(x_1\cdots x_{\mu})$ are obtained by summing $\pi_\rho(\pmb{x})$ over all unspecified bits, which, using Eq.~\eqref{eq:dpp_1}, reduces to a determinant formula involving the covariance matrix and is given by~\cite{Launay_2020,Kulesza_2012,Collura_2025} 
\begin{equation}\label{eq:formdet}
   \pi_{\rho}(x_1\cdots x_{\mu}) = \frac{\det[(\mathbb{1}_{[\mu+1,2L]} \,  + \, \mathbb{\Gamma})|_{(x_1\cdots x_{\mu},1...1)}]}{\det[\mathbb{1} + \mathbb{\Gamma}]}.
\end{equation}

Let us finally remark that this approach can be easily extended to any contiguous subsystem of qubits starting from the first lattice site within the full system. Indeed, any Pauli substring supported on a connected subsystem $[1,\ell]$ maps to a Majorana substring defined on the corresponding Majorana sublattice $[1,2\ell]$. Moreover, since the reduced density matrix of a Gaussian ensemble remains Gaussian, the same formula for the marginal probabilities applies. One simply needs to evaluate Eq.~(\ref{eq:formdet}) using the covariance matrix restricted to the subsystem under consideration.

\section{Non-stabilizerness in Gaussian Dynamics}\label{sec:monitored_dynamics}

We employ the methods outlined in the previous sections to study the Hamiltonian dynamics of the SREs $M_{\alpha}$ in non-interacting fermionic models. Focusing on the full-state magic, we employ a quantum-trajectory unraveling to investigate the impact of projective measurements. These measurements compete with the unitary spreading of correlations, shaping the development of extensive (or sub-extensive) non-stabilizerness in the stationary state. 

We study two different non-interacting Hamiltonian dynamics: a particle-number-conserving chain of hopping fermions, and the quantum Ising chain, which instead lacks this symmetry.

The Section is organized as follows. The results for the two previously mentioned models are presented separately in dedicated subsections — subsections~\ref{sec:hopping} and \ref{sec:ising}. 
Each of them is structured to first discuss the so-called \textit{plain dynamics}, i.e., the unitary evolution in the absence of measurements, followed by a detailed analysis of the \textit{monitored dynamics}.

\subsection{Hopping fermions}\label{sec:hopping}

The first model we consider consists on spinless fermions hopping on a ring with periodic boundary conditions, described by the Hamiltonian
\begin{equation}\label{eq:H_XX}
    \hat{H} = -\frac{1}{2} \sum_{j=0}^{L-1} \left( \hat{c}_j^\dagger \hat{c}_{j+1} + \hat{c}_{j+1}^\dagger \hat{c}_j \right),
\end{equation}
which preserves the total particle number. The dynamics is Gaussian due to the quadratic form of $\hat{H}$, with the two-point correlation matrix
$\mathbb{C}_{ij}(t) = \langle \hat{c}_i^\dagger(t) \hat{c}_j(t) \rangle$ evolving as:
$\mathbb{C}(t+s) = \mathbb{R}^\dagger(s) \mathbb{C}(t) \mathbb{R}(s)$,
where $\mathbb{R}_{mn}(s)$ encodes the free evolution.
We focus on a protocol where we measure the local occupation numbers $\hat{n}_j = \hat{c}_j^\dagger \hat{c}_j$, which are quadratic operators. These measurements preserve the Gaussianity of the state due to the Wick theorem and the form of the projection operators. Specifically, for each time step $dt$, we consider a measurement process as in Ref.~\cite{Coppola_2022}: for site $k$, we perform a measurement of $\hat{n}_k$ with probability $\gamma dt$, where $\gamma$ defines the measurement rate.  If a measurement occurs, the outcome $n = 0,1$ is drawn with probability $p_k(n) = (1-n) + (2n-1) \langle \hat n_k \rangle$, and the two-point function is updated consequently. \\

\begin{figure}[t!]
\includegraphics[width=0.5\textwidth]{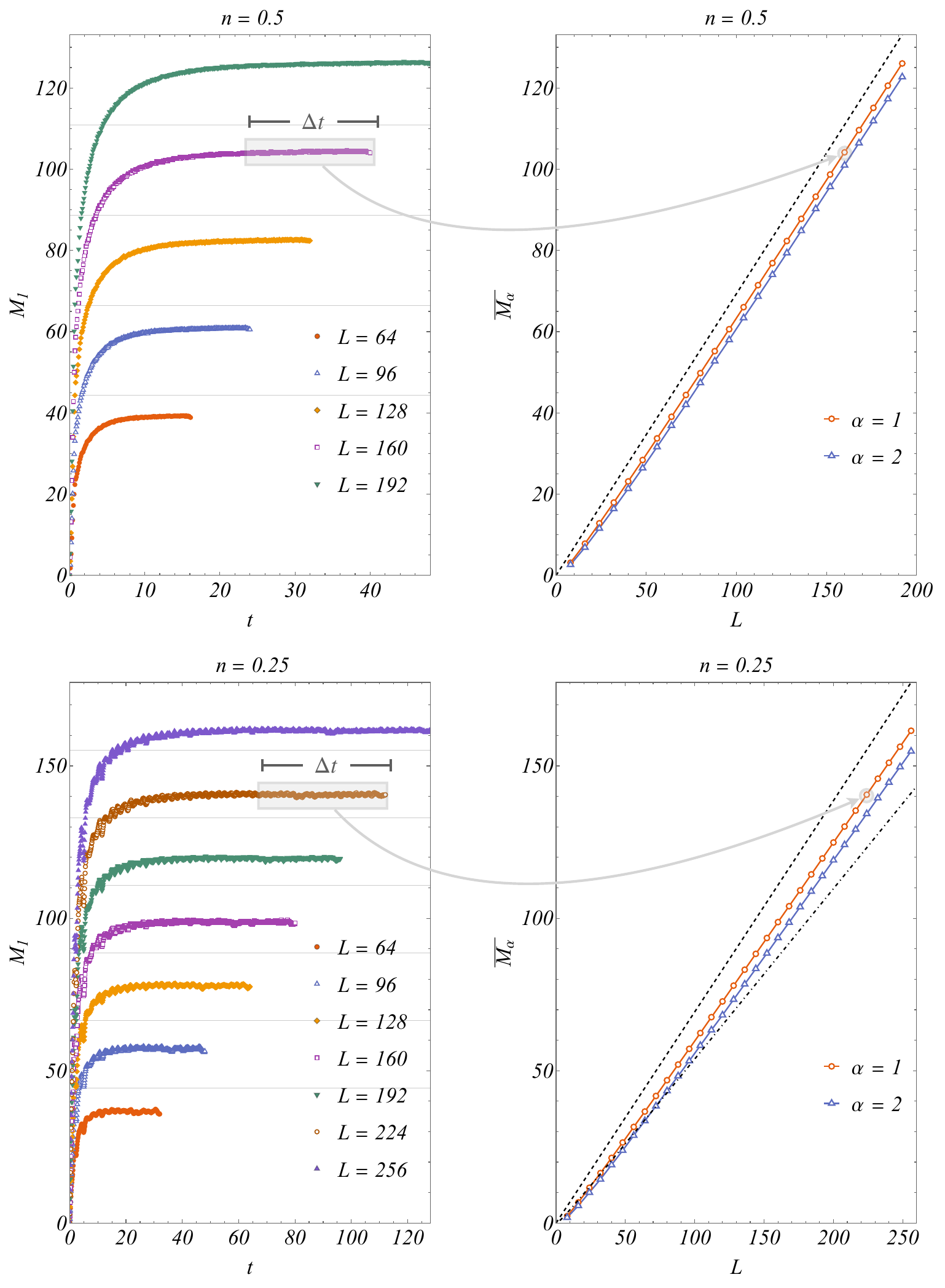}
\caption{\label{fig:XX_plain}\textbf{Left panels}. Plain time evolution (no measurement, i.e. $\gamma=0$) under the hopping Hamiltonian in Eq.~(\ref{eq:H_XX}) of the SRE $M_1$ after having initialized the system into the N\'eel state with particle density $n = N/L = 1/2$ (top) and $n = N/L = 1/4$ (bottom). Each point has been obtained by averaging over $\mathcal{S}=1000$ Pauli strings. Gray horizontal lines correspond to $L\log(2)$. \textbf{Right panels}. Extensive behavior of the stationary stabilizer entropies averaged over a time window $\Delta t = [L/8,L/4]$ for $n=1/2$, and $\Delta t = [L/4,L/2]$ for $n=1/4$. The dashed line corresponds to $L \log(2)$ while, for $n=1/4$ also $\log\binom{L}{L/4}$ is drawn (dot-dashed line).}
\end{figure}

\paragraph{Plain dynamics. ---}
We begin our analysis by considering the purely unitary dynamics generated by the hopping Hamiltonian: the dynamics is fully deterministic, and the presence of a $U(1)$ symmetry in the model can significantly influence the stationary value of the non-stabilizerness reached at late times.

We consider two different initial conditions corresponding to distinct particle-number sectors, namely $N = L/2$ and $N = L/4$.
These are prepared as generalized Néel states with fixed particle densities, i.e., $\ket{1010\cdots}$ and $\ket{10001000\cdots}$, respectively.
The SREs $M_\alpha$ increase over time and eventually reach stationary values that scale linearly with the system size. The data are subject to fluctuations due to the probabilistic technique used in computing the SREs. In particular, they have been computed by averaging over $\mathcal{S}=1000$ samples. 
Notably, after performing a time average over a suitable window to suppress finite-time fluctuations (see Fig.~\ref{fig:XX_plain}, where we show the typical dynamics of $M_1$ for different system sizes), we observe that the stationary value of the magic exhibits, on top of the expected extensive behavior, sub-leading logarithmic corrections such that
$
\overline{M_{\alpha}(L)} \sim a_{\alpha}L - b_{\alpha} \log L - c_{\alpha}
$.
Here the coefficients ${a_{\alpha}, b_{\alpha}, c_{\alpha}}$ generally depend on both the Rényi index $\alpha$ and the initial conditions, i.e. the fixed particle density $n = N/L$ that characterizes the dynamical sector. For notational simplicity, we omit below the explicit dependence on $n$.

In Fig.~\ref{fig:XX_plain}, we display the leading behavior of the stationary values of the SREs for the two initial particle densities and for Rényi indices $\alpha = 1, 2$. At half filling, the non-stabilizerness exhibits extensive behavior that asymptotically approaches the Haar average value of 
$L\log(2)$ In the large-$L$ limit, the curves become nearly parallel to this reference, indicating convergence. This agreement becomes even more apparent when examining the subleading logarithmic corrections. This behavior aligns with what has been observed for random Gaussian states~\cite{Collura_2025}, even though is not clear which role is playing the $U(1)$ symmetry.
In contrast, for the case with particle number $N = L/4$, the SREs do not appear to relax to the Haar value associated with the full Hilbert space. This may still occur, but, as expected for systems initialized with lower particle densities, reaching the thermodynamic limit likely requires significantly larger system sizes and longer times. Nonetheless, the SREs also deviate from the bound set by the Hilbert space dimension within the fixed-particle-number sector. In this case, the maximal scaling would be proportional to $\log \binom{L}{L/4}$, which disagrees with our numerical findings, as the numerics actually exceed this value. We find this interesting, as to the best of our knowledge, the non-stabilizerness uniformly averaged over a fixed particle-number sector is expected to be bounded from above by the logarithm of its dimension~\cite{Russomanno_2025}.

\begin{figure}[t!]
\includegraphics[width=0.5\textwidth]{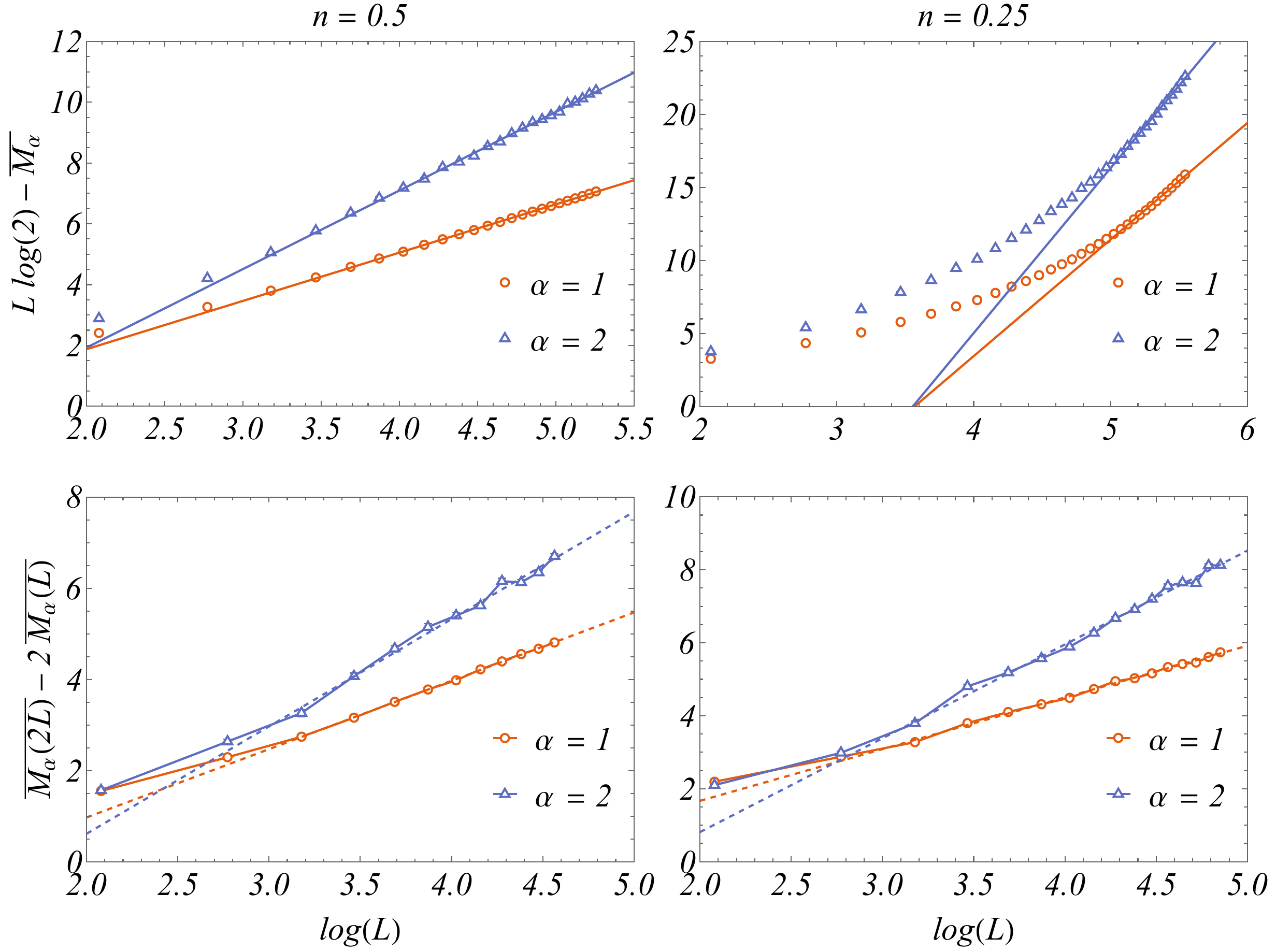}
\caption{\label{fig:XX_plain_log}\textbf{Top panels}. Late-time stabilizer Rényi entropies after a plain dynamics in the hopping fermions for $n = 1/2$ and $n = 1/4$, after subtracting the (non-symmetric) Haar average $L \log(2)$.  
\textbf{Bottom panels}. Logarithmic corrections to the non-stabilizerness, extracted via finite-size analysis (see main text for details).
}
\end{figure}

As already mentioned, the SREs are expected to exhibit sub-leading logarithmic corrections to the leading extensive behavior. In Fig.~\ref{fig:XX_plain_log}, we report a detailed analysis for both cases. As a first naive approach, we subtract from the stationary entropies the maximal Haar value, $L \log(2)$, and plot the residuals on a logarithmic scale, disregarding the presence of conserved quantities constraining the dynamics of non-stabilizerness.
Remarkably, for the half-filling case $n = 1/2$, the remaining corrections display a striking logarithmic growth. This unambiguously confirms that the extensive coefficient of quantum magic under plain unitary dynamics, starting from a canonical Néel state, indeed coincides with the Haar value. 
The quarter-filling case $n = 1/4$ shows a different behavior. After subtracting the Haar average, no clear logarithmic trend is observed, suggesting that the Haar value is not the appropriate leading term to subtract in this case.

To further investigate the logarithmic subleading corrections, without relying on any assumed leading extensive contribution, we analyze the finite-size difference $\overline{M_{\alpha}(2L)} - 2 \overline{M_{\alpha}(L)}$. 
This combination cancels out any extensive linear scaling, allowing us to isolate potential logarithmic terms (see the bottom panels of Fig.~\ref{fig:XX_plain_log}).

This refined analysis enables us to extract the logarithmic coefficients for both particle densities and Rényi indices. Interestingly, the logarithmic corrections appear to be weakly dependent on the particle density, yet they depend on the Rényi index $\alpha$. Our best-fit estimates yield the following coefficients:
\begin{align*}
n=1/2, \; b_{1} &= 1.50(1), \; b_{2} = 2.33(9), \\
n=1/4, \; b_{1} &= 1.42(2), \; b_{2} = 2.57(6).
\end{align*}

\begin{figure}[t!]
\includegraphics[width=0.5\textwidth]{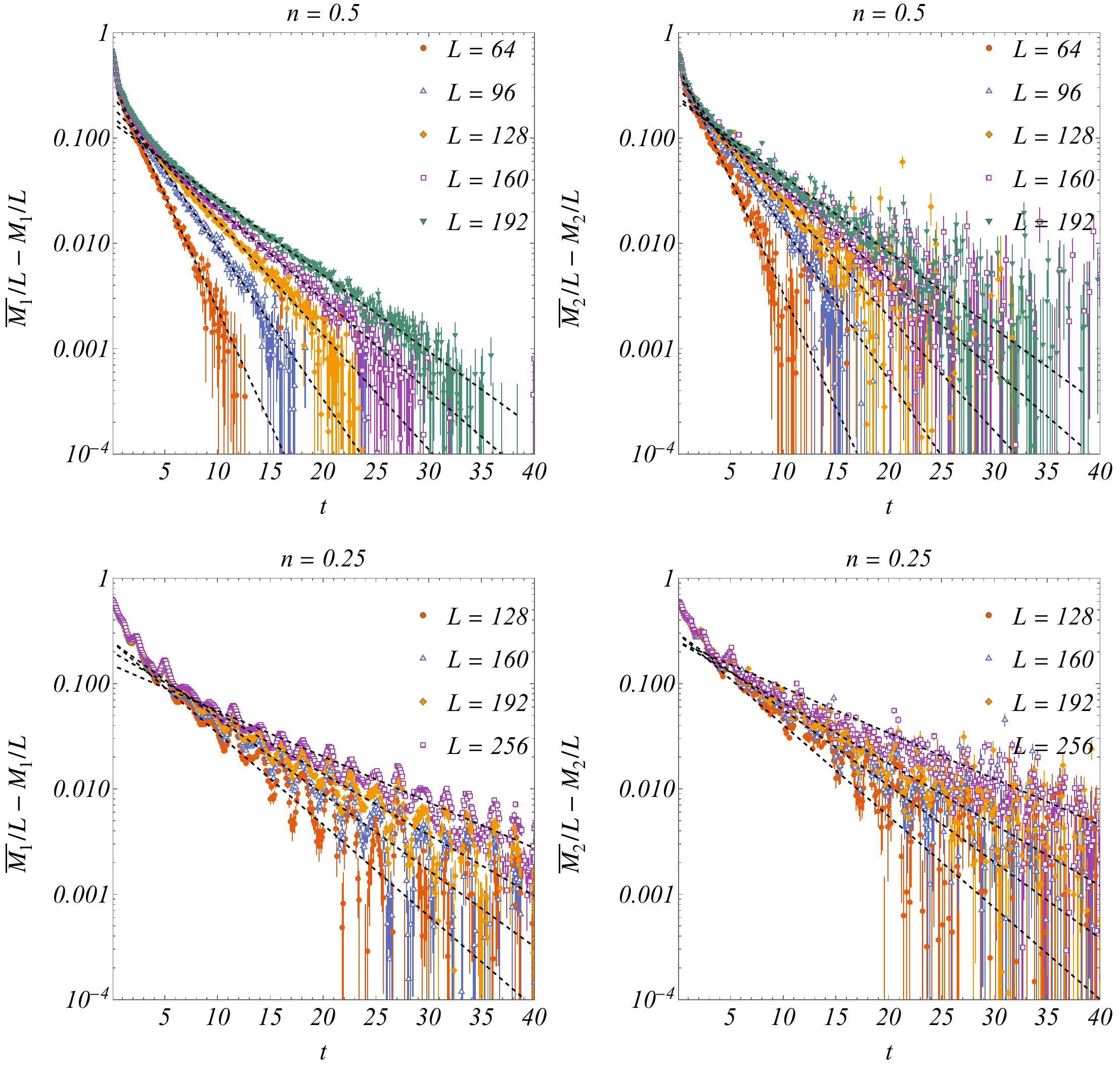}
\caption{\label{fig:XX_plain_exp} Log-linear plot of the relaxation dynamics of the SREs after having subtracted the best-fit finite-size stationary values $\overline{M_{\alpha}(L)} \sim a_{\alpha}L - b_{\alpha} \log L - c_{\alpha}$; dashed lines are size-dependent exponential decay drown as guide for the eyes.}
\end{figure}

We conclude our discussion of the plain unitary dynamics of hopping fermions by examining the time evolution of non-stabilizerness itself, i.e., how the SREs relax toward their stationary values. While our previous analysis focused on the asymptotic behavior, it is equally insightful to investigate the dynamical approach to stationarity. To this end, we consider the SRE density $M_{\alpha}(t,L)/L$, which depends on both time and system size, and focus on its behavior at late times, $t \gg 1$.
The limits $L \to \infty$ and $t \to \infty$ do not necessarily commute. In this section, although we analyze both the $n = 1/2$ and $n = 1/4$ cases, as well as the two Rényi indices $\alpha = 1, 2$, it is evident that the data are significantly more accurate for $n = 1/2$. This is due to reduced fluctuations and, most importantly, the necessity of knowing the extensive contribution to the SREs in the stationary regime with high precision. Consequently, one must exercise caution when interpreting relaxation dynamics in finite systems—especially when aiming to infer asymptotic scaling behavior.

The first question that naturally emerges is how the SREs approach their asymptotic form at large times. As a purely finite-size effect, and in line with observations from both random circuit dynamics and Gaussian evolutions~\cite{Turkeshi_2024_2,Collura_2025}, the approach to stationarity may follow an exponential decay. 
As a preliminary analysis, we thus consider how the time-dependent data approach the finite-size stationary values. These stationary values have been carefully extracted from our best-fit analysis, which accounts for the extensive ($\sim L$), logarithmic ($\sim\log L$), and decaying ($\sim L^{-1}$) corrections, as already discussed. To this end, we study the quantity $(\overline{M_{\alpha}(L)} - M_{\alpha}(t, L))/L$ (see Fig.~\ref{fig:XX_plain_exp}). This difference exhibits an exponential decay toward the finite-size stationary values, indicating that the $t \to \infty$ limit is taken first in this analysis. However, as the system size increases, the curves begin to bend on the log-linear scale and increasingly overlap in the early-time regime. This signals the emergence of a different universal behavior associated with the thermodynamic limit. This feature is particularly evident for the $\alpha = 1$, $n = 1/2$ case, where the data are especially clean and reliable.

\begin{figure}[t!]
\includegraphics[width=0.5\textwidth]{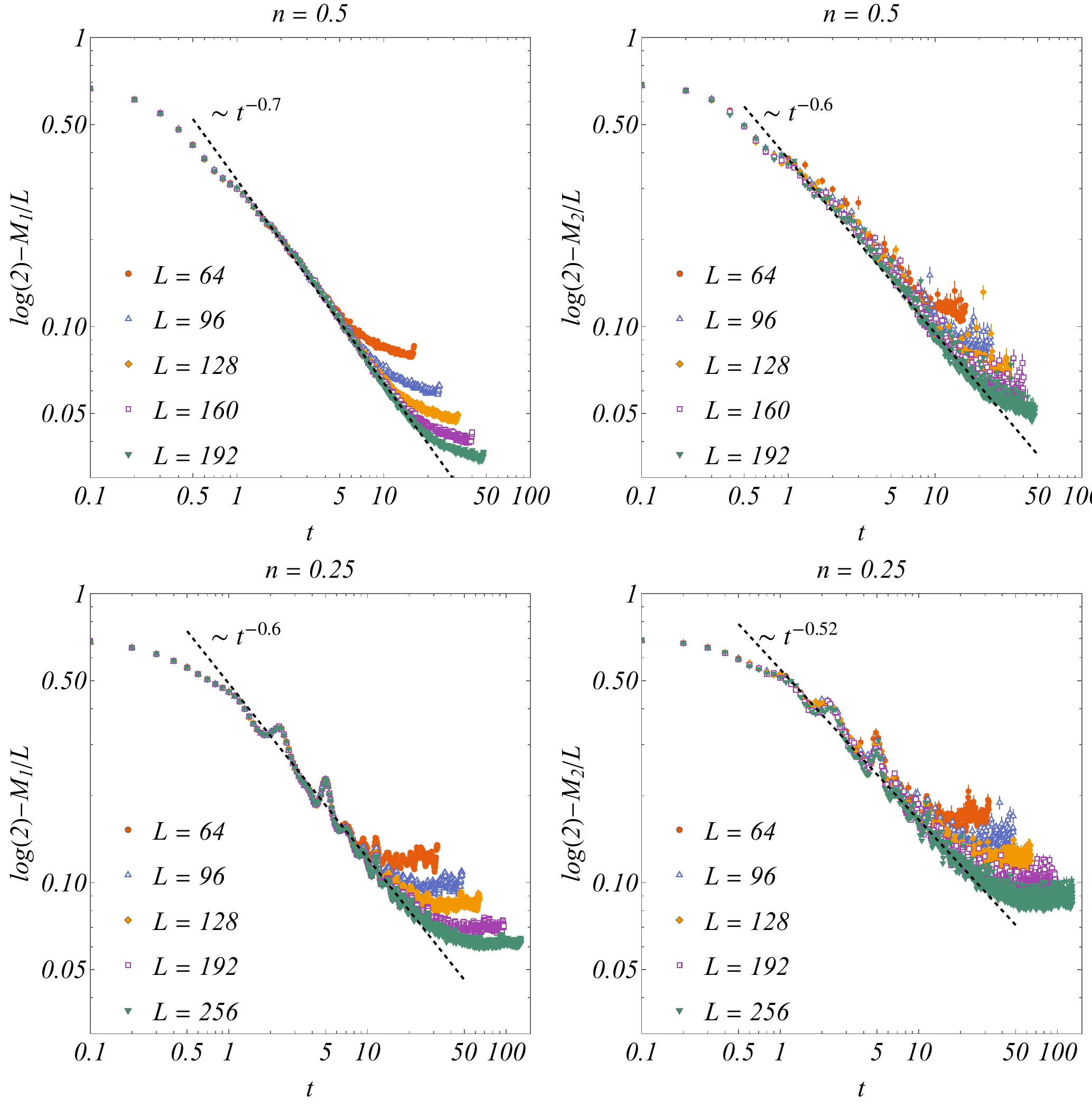}
\caption{\label{fig:XX_plain_pow} Time evolution of the stabilizer Rényi entropy density approaching its thermodynamic limit for quenches with the Hopping Hamiltonian. The log-log plot highlights the thermodynamic scaling regime for times $t < t^*(L)$, where finite-size effects are negligible. The dashed line indicates the algebraic decay observed in this regime, as discussed in the main text.
}
\end{figure}

A more systematic and conceptually sound analysis requires taking the thermodynamic limit ($L \to \infty$) before the long-time limit ($t \to \infty$). In this regime, the time-dependent behavior should become independent of $L$, and data from different system sizes are expected to collapse onto a universal thermodynamic scaling curve. As shown in Fig.~\ref{fig:XX_plain_pow}, after shifting the data by a constant offset $\log(2)$ the SRE density $M_\alpha(t,L)/L$ displays an algebraic relaxation at intermediate times. However, once finite-size effects become relevant, i.e., beyond a characteristic crossover time $t^*(L)$, the dynamics transitions to the previously discussed exponential approach to the finite-size stationary value. Importantly, this exponential regime is delayed to progressively later times as the system size increases, and in the thermodynamic limit, it ultimately disappears, leaving the algebraic decay as the dominant relaxation behavior.

A complementary analysis of the stationary values of the subsystem stabilizer Rényi entropies, based on the Generalized Gibbs Ensemble description, is presented in Appendix~\ref{sec:appendix_GGE}. \\

\paragraph{Monitored dynamics. ---}

In the following, we investigate the unitary hopping dynamics interspersed with local projective measurements of the occupation number, performed at a finite measurement rate $\gamma \neq 0$. Our analysis is restricted to the half-filling sector with particle density $n = 1/2$ where the system is initialized in the N\'eel state.

For each value of the system size $L$ and measurement rate $\gamma$, we have collected $N_{\mathrm{traj}} = 500$ independent quantum trajectories, each corresponding to a distinct random realization of local projective measurements. For each trajectory, the stabilizer R\'enyi entropies are evaluated via perfect sampling, using at least $\mathcal{S} \in [1000,2000]$ configurations of Majorana string operators depending on the system size.

\begin{figure}[t!]
\includegraphics[width=0.5\textwidth]{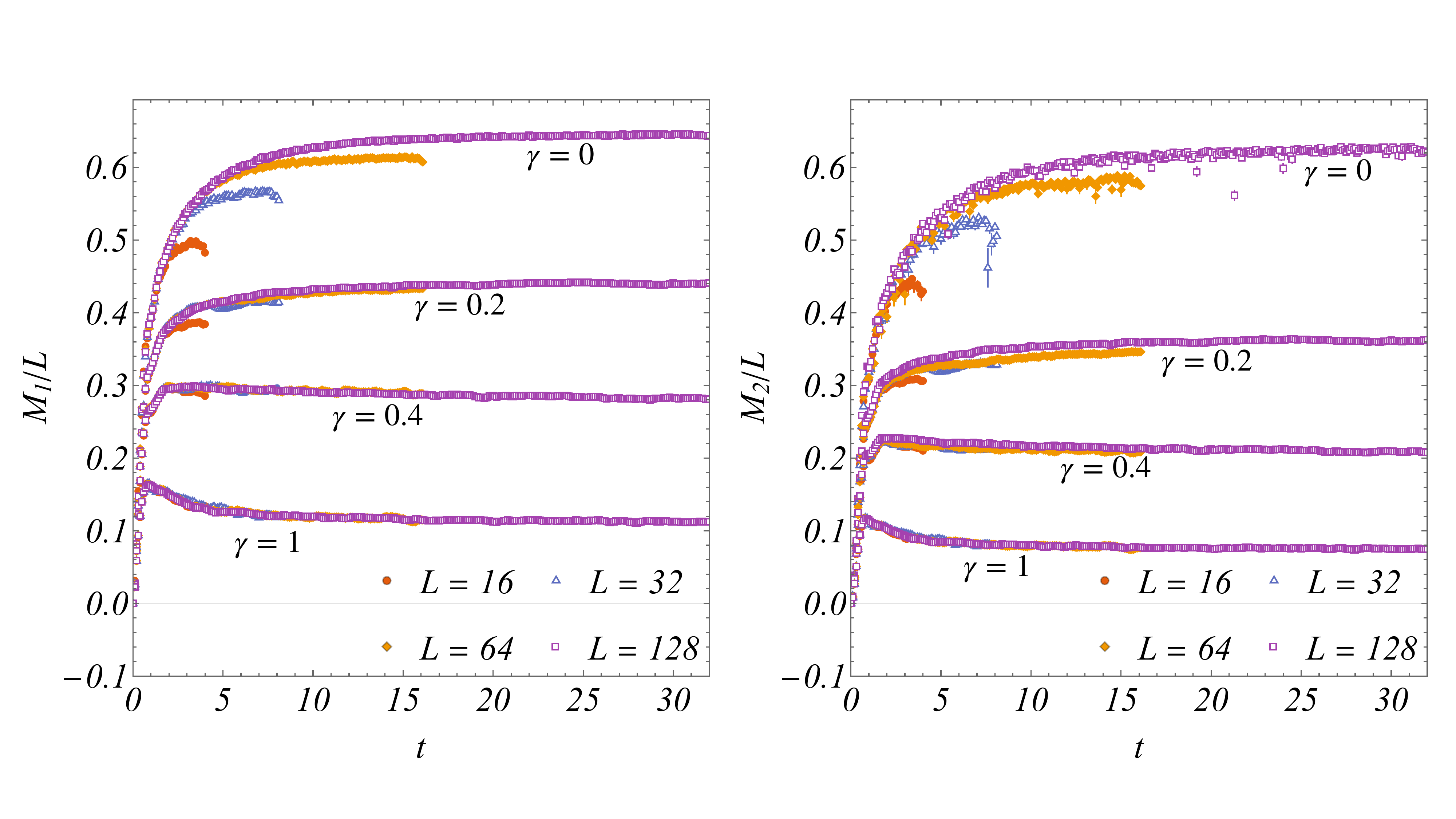}
\caption{\label{fig:XX_neel_monitor} Time evolution of the stabilizer Rényi entropy densities $M_1/L$ (left panel) and $M_2/L$ (right panel) under unitary free-fermion dynamics interspersed with local projective measurements performed at various rates $\gamma$. The system is initialized in the N\'eel state at half filling ($n = 1/2$), and averages are taken over $N_{\mathrm{traj}} = 500$ independent quantum trajectories. The rescaled entropies exhibit an overall extensive behavior, with finite-size corrections becoming increasingly suppressed as the measurement rate $\gamma$ increases.}
\end{figure}

Our first analysis focuses on the impact of a finite measurement rate on the time-dependent behavior of the SREs. In particular, Fig.~\ref{fig:XX_neel_monitor} illustrates the dynamics of non-stabilizerness for system sizes $L = 16, 32, 64, 128$ and for some representative values of the measurement rate $\gamma \in [0,1]$. As expected, once the SREs are rescaled to their extensive component (i.e., plotting their density) the overall qualitative behavior remains robust across different measurement regimes. Despite significant variation in $\gamma$, the relaxation towards the stationary value appears to retain an algebraic character, at least qualitatively.

What is particularly notable is the behavior of the subleading finite-size corrections, visible as deviations from the thermodynamic-limit curves. These are more pronounced for small values of $\gamma$, but decrease and eventually become negligible as the measurement rate increases. This suggests that larger measurement rates suppress the features associated with finite-size effects, promoting faster convergence to the thermodynamic limit.

To more accurately characterize the leading behavior of non-stabilizerness in the stationary regime, we follow an approach analogous to that used in the case of purely unitary dynamics. Specifically, we compute the time average of the stabilizer Rényi entropies over a suitable late-time window, and then analyze the scaling of these averaged values as a function of the total system size $L$.

\begin{figure}[t!]
\includegraphics[width=0.5\textwidth]{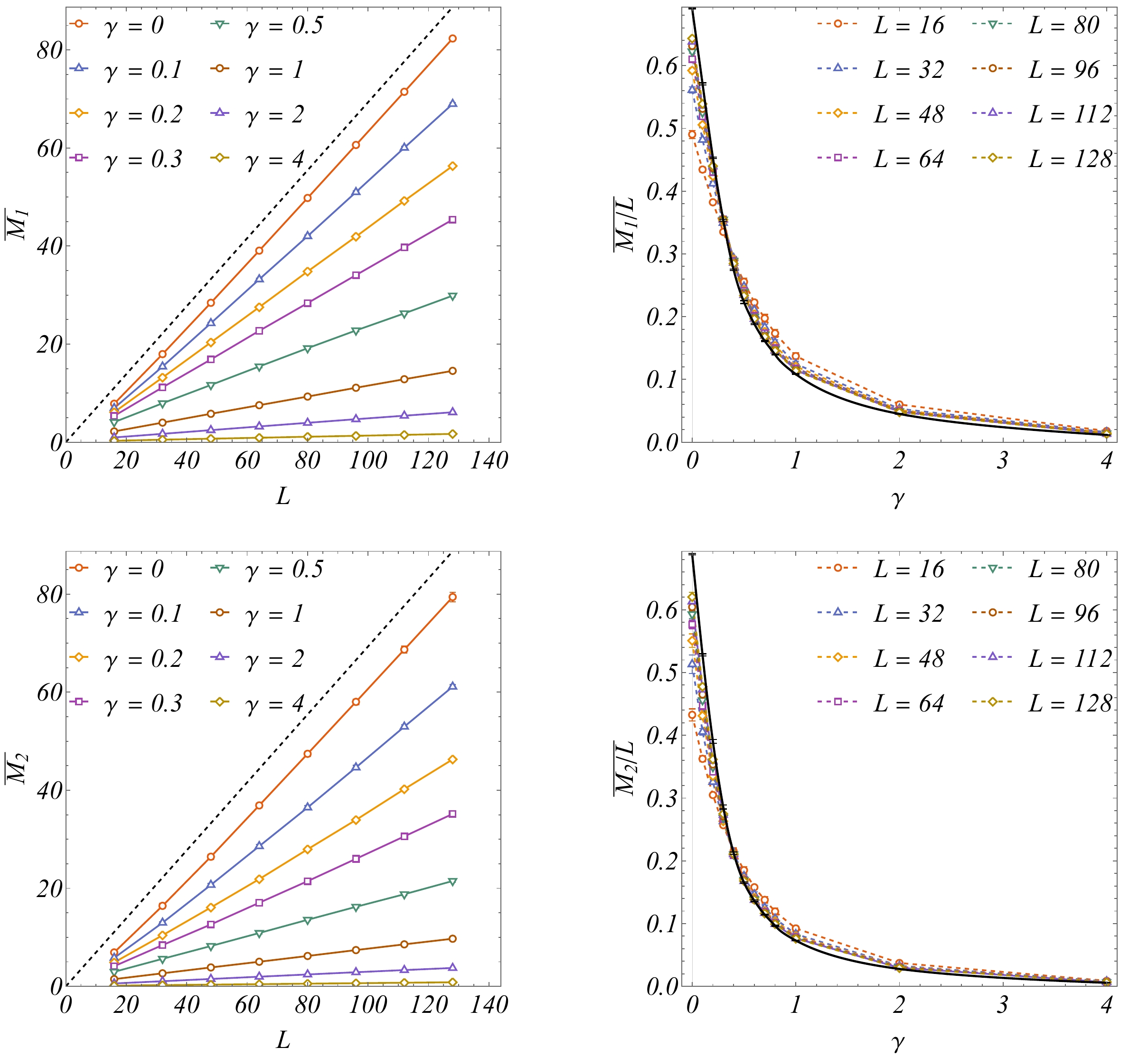}
\caption{\label{fig:XX_neel_monitor_scaling} Scaling of the stabilizer Rényi entropy densities $M_1/L$ and $M_2/L$ in the stationary regime for various values of the measurement rate $\gamma$. In each panel, the data points represent time-averaged values obtained from average time series (cfr. Fig.~\ref{fig:XX_neel_monitor}), computed over a suitable late-time window. \textbf{Left panels}. The entropy densities as functions of the system size $L$, highlighting their extensive scaling behavior (the dashed black line is $L \log(2)$). \textbf{Right panels}. The extracted leading coefficients (i.e., slopes from linear fits in $L \in [16,128]$) as functions of $\gamma$ (full black lines), demonstrating that the non-stabilizerness density varies smoothly with the measurement rate without exhibiting abrupt transitions.
}
\end{figure}

The raw time-averaged data are shown in the left panels of Fig.~\ref{fig:XX_neel_monitor_scaling}, plotted as a function of the system size for several representative values of the measurement rate $\gamma$. The same data are displayed in the right panels as a function of $\gamma$, for various fixed system sizes. To extract the leading extensive contribution, we perform a linear fit in the range $L \in [16, 128]$. The resulting slope, which corresponds to the density of non-stabilizerness, is indicated by a solid black line in the right panels of Fig.~\ref{fig:XX_neel_monitor_scaling}.

Our results confirm that the extensive behavior of non-stabilizerness persists under projective measurements and evolves smoothly as a function of the measurement rate $\gamma$. Notably, no abrupt transition is observed in this scaling behavior, suggesting a continuous crossover rather than a sharp dynamical phase transition.

\begin{figure}[t!]
\includegraphics[width=0.5\textwidth]{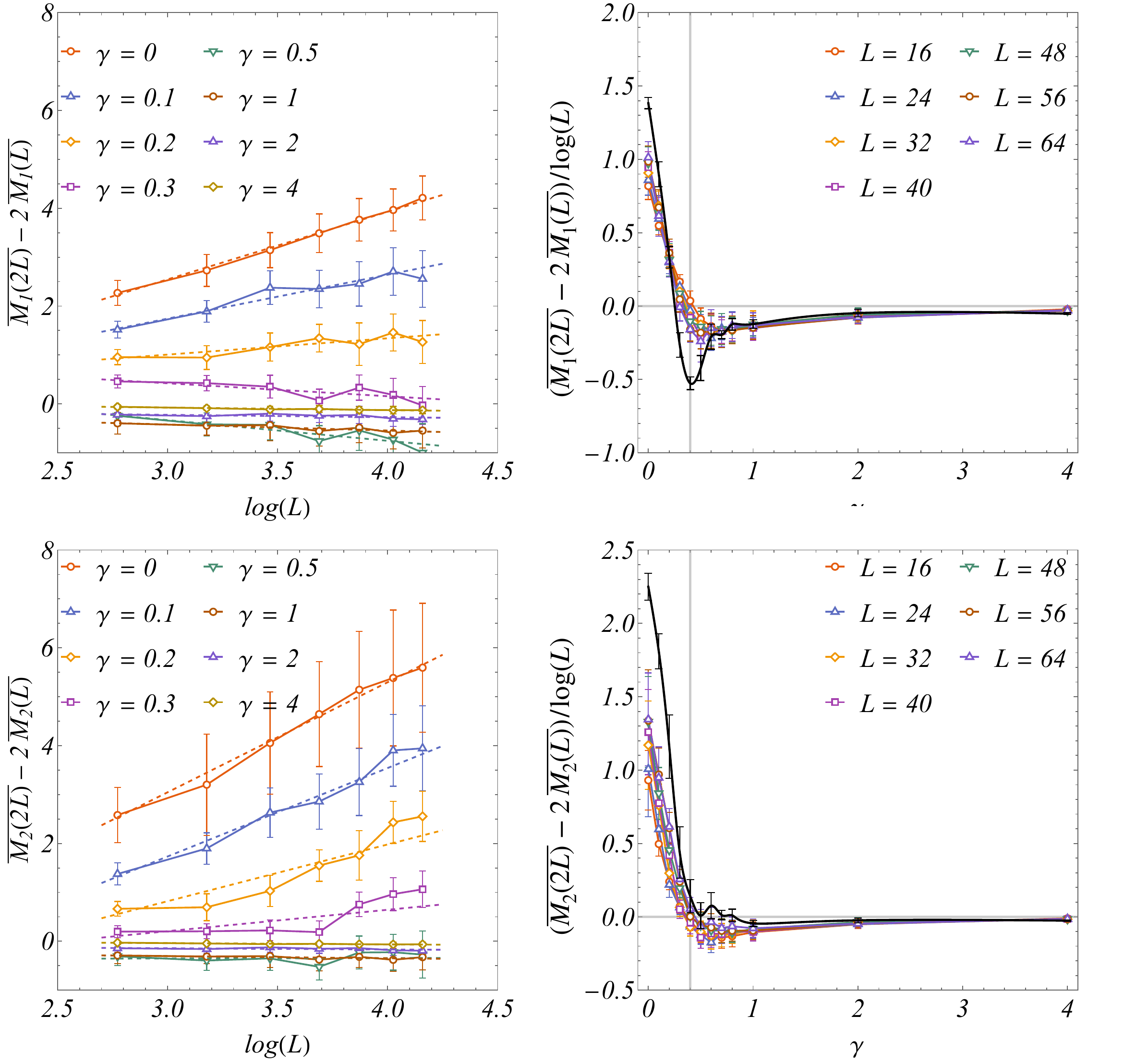}
\caption{\label{fig:XX_neel_monitor_log} \textbf{Left panels}. Finite-size difference $\overline{M_{\alpha}(2L)} - 2 \, \overline{M_{\alpha}(L)}$ as a function of $\log(L)$ for various values of the measurement rate $\gamma$. Dashed lines indicate the best linear fits, from which the logarithmic coefficients are extracted. \textbf{Right panels}. The same data, rescaled by $\log(L)$, are shown as a function of $\gamma$. Solid black lines denote the extrapolated values of the coefficient of the logarithmic corrections. Vertical gray lines mark the threshold $\gamma^{*} \sim 0.4$, above which the numerical data become statistically compatible with zero within $2\sigma$.}
\end{figure}

On the contrary, as already hinted at the level of the time-dependent data, and in line with recent observations on canonical participation entropies~\cite{Sierant_2022}, the stationary quantum magic appears to exhibit an abrupt transition in its subleading corrections (here logarithmic), rather than in its leading extensive behavior. To further investigate this feature we once again analyze the finite-size difference $\overline{M_{\alpha}(2L)} - 2 \, \overline{M_{\alpha}(L)}$,
as done in the unitary case. We study the behavior of this quantity as a function of the measurement rate $\gamma$, shown in Fig.~\ref{fig:XX_neel_monitor_log}. From the numerical data, we extract the slope, which corresponds to the coefficient of the logarithmic correction in the thermodynamic limit. Although our system sizes are limited to $L \leq 128$, the results allow us to clearly identify a qualitative change in this coefficient as the measurement rate increases beyond a threshold value $\gamma^* \sim 0.4$. Specifically, upon rescaling the finite-size differences by $\log(L)$ and plotting them as a function of $\gamma$, the data become statistically compatible with zero above this threshold. This observation strongly indicates that the effect of measurements manifests itself predominantly in the sub-leading corrections to the non-stabilizerness of the monitored non-interacting hopping fermions, with a sharp suppression of the logarithmic contribution once $\gamma \gtrsim \gamma^*$.

\subsection{Ising quantum chain}\label{sec:ising}
The second scenario consists of a quantum spin chain governed by the transverse field Ising model (TFIM) with periodic boundary conditions. The Hamiltonian reads:
\begin{equation}\label{eq:H_TFIM}
    \hat{H} = -\sum_{j=0}^{L-1} \left( \hat{\sigma}_j^x \hat{\sigma}_{j+1}^x + h \, \hat{\sigma}_j^z \right),
\end{equation}
where $\hat{\sigma}_j^\alpha$ ($\alpha = x, y, z$) are Pauli matrices acting on site \(j\), and \(h\) is the transverse magnetic field. 

Although formulated in terms of spin operators, the model can be exactly mapped to a system of free fermions through the Jordan-Wigner transformation, and further reformulated in terms of Majorana fermions. 

The dynamics is Gaussian and fully encoded by the Majorana covariance matrix $\mathbb{\Gamma}_{\mu\nu}$ (as introduced in Section~\ref{sec:gaussian}). 
Equivalently, one can consider the Majorana two-point function
$\mathbb{G}_{\mu\nu} = \langle \hat{\gamma}_\mu \hat{\gamma}_\nu \rangle$,
from which the covariance matrix is recovered as:
$\mathbb{\Gamma} = -i\left(\mathbb{G} - \mathbb{1}\right)$.

The time evolution of the two-point function is governed by a real orthogonal transformation  $\mathbb{R}(t) \in \mathrm{SO}(2L)$ as
$\mathbb{G}(t + s) = \mathbb{R}(s)\, \mathbb{G}(t)\, \mathbb{R}^{T}(s)$, reflecting the quadratic nature of the dynamics in the Majorana basis.

\begin{figure}[t!]
\includegraphics[width=0.5\textwidth]{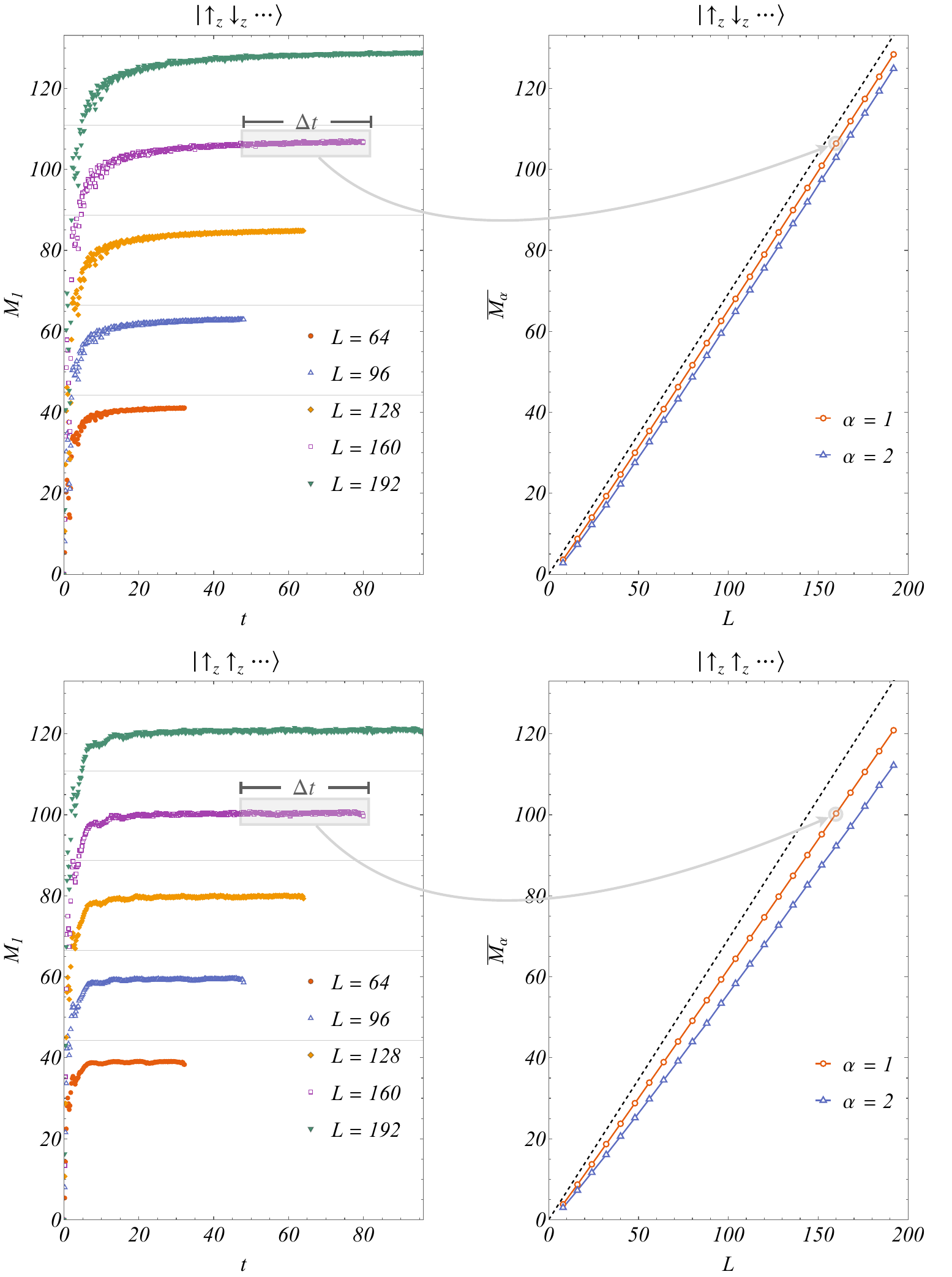}
\caption{\label{fig:ising_plain}\textbf{Left panels}. Plain time evolution (no measurement) under the Ising Hamiltonian with $h=0.5$ in Eq.~(\ref{eq:H_TFIM}) of the stabilizer Renyi entropy $M_1$ after having initialized the system in the Néel state (top), or in the fully polarised state (bottom). Each point has been obtained by averaging over $\mathcal{S}\in[1000,24000]$ sample of Pauli strings, depending on the system size. Gray horizontal lines correspond to $L\log 2$. \textbf{Right panels}. Extensive behavior of the stationary stabilizer entropies averaged over a time window $\Delta t = [L/4,L/2]$. The dashed line corresponds to $L \log2$.}
\end{figure}

We initialize the system in a fully polarized state along the $z$-direction
$|\!\uparrow_z \cdots \uparrow_z\rangle$, which in the fermionic language represents the vacuum state, or in the N\'eel state $|\!\uparrow_z \downarrow_z\uparrow_z \downarrow_z \cdots \rangle$ which corresponds to $\ket{0101\cdots}$.
Both initial states are Gaussian in terms of Majorana fermions, and they are thus completely specified by the two-point correlation function. The unitary dynamics generated by $\hat H$ preserves the Gaussianity and the structure of the correlation matrix.  
In our setup, the system undergoes stochastic projective measurements in the $\hat{\sigma}^z_k$ basis with a characteristic rate $\gamma$. These measurements can be implemented in two complementary protocols. In the first, local $\hat{\sigma}^z_k$ observables are measured probabilistically with Poissonian statistics and average waiting time $\gamma$ per site, 
with the state updated according to the Born rule~\cite{Tirrito_2023}.
In the second, we consider the quantum jump formalism, where measurements act abruptly and are described by a stochastic Schrödinger equation~\cite{Paviglianiti_2023}. The resulting evolution, particularly in the no-click limit, is governed by an effective non-Hermitian Hamiltonian. This provides a distinct perspective on the influence of weak monitoring, capturing different dynamical regimes depending on the frequency and nature of the measurement events.
Importantly, both protocol preserve the Gaussian character of the state. Therefore, the system at all times is fully described by its two-point Majorana correlation matrix, and higher-order correlations can be reconstructed via Wick's theorem. \\

\paragraph{Plain dynamics. ---}

Here we consider a quench dynamics generated by the TFIM deeply within its ferromagnetic phase; specifically, we choose $h = 0.5$. In this setting, no trajectory averaging is involved, as no stochastic measurement events occur. 

Also here, the stabilizer Rényi entropies $M_{\alpha}$ increase over time and eventually saturate to stationary values that scale linearly with the system size. The numerical data exhibit fluctuations arising from the probabilistic nature of the computation, as the SREs are estimated via sampling. In particular, the averages are taken over $\mathcal{S} \in [1000, 24000]$ Majorana string configurations, depending on the system size.

In Figure~\ref{fig:ising_plain}, we present representative time-series data for the stabilizer Rényi entropy with index $\alpha = 1$. After averaging the data over a suitable late-time window, we extract the leading extensive behavior as a function of system size. When the dynamics starts from the N\'eel state, the stationary values approach the expected Haar limit $L \log 2$ at leading order, with subleading deviations similar to those observed in Fig.~\ref{fig:XX_plain}. Conversely, when starting from the vacuum state the system appears to equilibrate more rapidly in time. However, the subleading corrections to the extensive scaling can be more pronounced in this case, indicating that larger system sizes are necessary to unambiguously observe convergence toward the Haar average.

\begin{figure}[t!]
\includegraphics[width=0.5\textwidth]{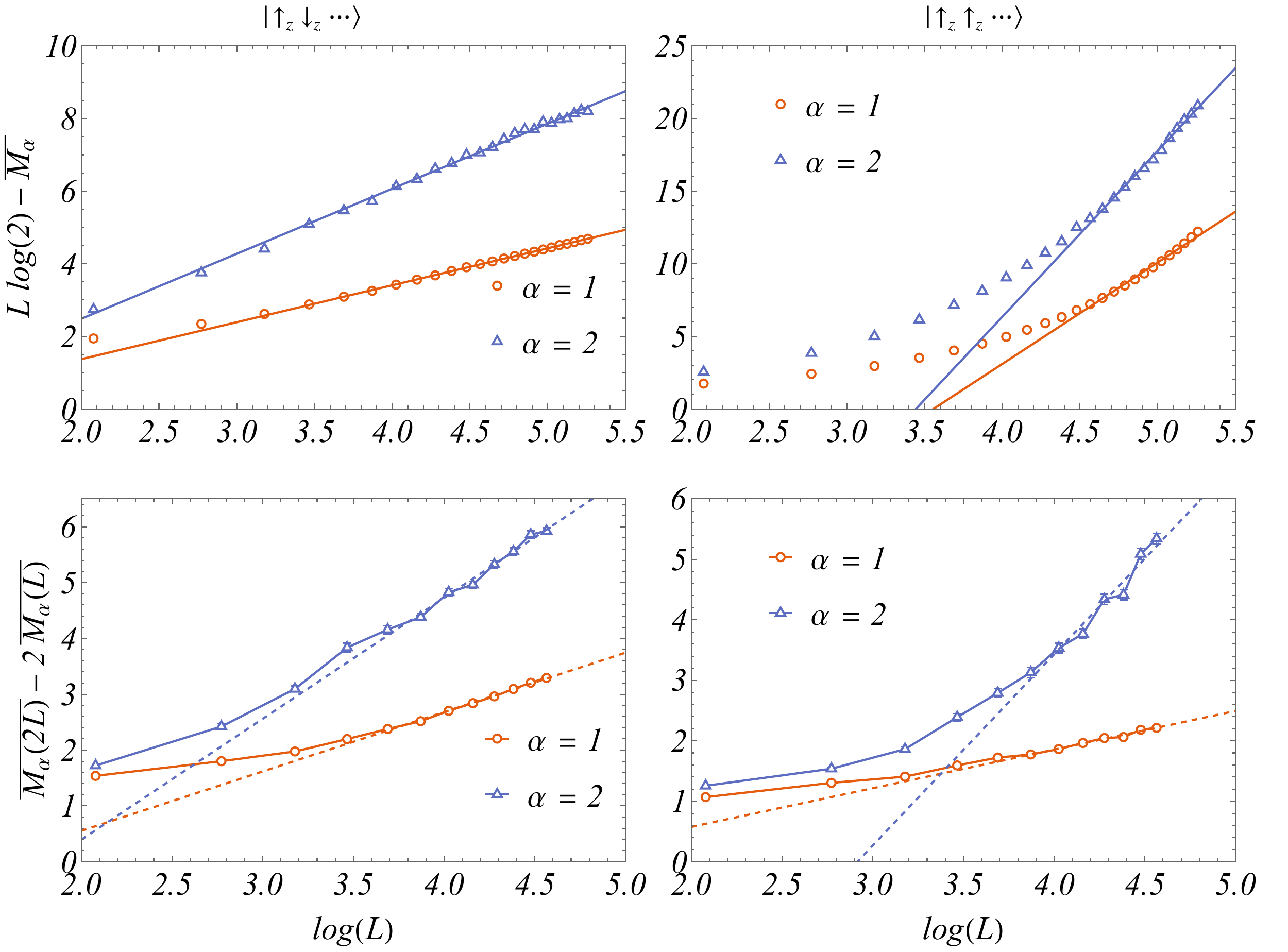}
\caption{\label{fig:ising_plain_log}\textbf{Top panels}. Late-time stabilizer Rényi entropies induced by a quench dynamics (without measures) in the Ising Hamiltonian with $h=0.5$, starting from the N\'eel and the vacuum initial states, after having subtracted the Haar average $L \log 2$.  
\textbf{Bottom panels}. Logarithmic corrections to the non-stabilizerness, obtained via finite-size analysis (see main text for details).
}
\end{figure}

Since we expect that $\overline{M_{\alpha}(L)} \sim a_{\alpha}L - b_{\alpha} \log L - c_{\alpha}
$ also in this setup, as found for hopping fermions, we analyze the subleading logarithmic behavior. To this end, we subtract the maximal Haar value from the stationary SREs and plot the residuals on a logarithmic scale. The results reveal a strikingly clean logarithmic correction to the leading term in the case of the Néel state (Figure~\ref{fig:ising_plain_log}). In contrast, the results for the vacuum initial state display deviations from the expected Haar-like behavior. Given the absence of $U(1)$ symmetry, one might anticipate that the leading extensive scaling of the stabilizer Rényi entropies follows the Haar form. However, significant finite-size effects are present: as shown in Fig.~\ref{fig:ising_plain}, the curves only begin to exhibit a parallel trend with the Haar prediction for very large system sizes. While this suggests a possible convergence toward the Haar slope, we cannot exclude the possibility that the true extensive coefficient in the thermodynamic limit differs from the Haar value.

To further probe this subleading structure, we once again examine the finite-size difference
$
\overline{M_{\alpha}(2L)} - 2\, \overline{M_{\alpha}(L)}
$
to isolate possible logarithmic terms. The results are displayed in the bottom panels of Figure~\ref{fig:ising_plain_log}.
This refined analysis enables us to extract the logarithmic coefficients, whose best-fit estimates yield
\begin{align*}
\textrm{N\'eel state:  }  \; \; b_{1} &= 1.06(3), \; b_{2} = 2.17(11), \\
\textrm{Vacuum state:  } \;  \;b_{1} &= 0.64(3), \; b_{2} = 3.16(26) \;.
\end{align*}

Also in this case, the logarithmic corrections exhibit a clear dependence on the Rényi index $\alpha$. Moreover, we observe a residual dependence on the choice of the initial state. Although such dependence might seem unexpected since we are probing the stationary properties of non-stabilizerness as a function of system size, and no explicit symmetry constraints the dynamics, this behavior can be attributed to finite-time effects and to the structure of the integrals of motion. Indeed, while the model lacks a $U(1)$ conservation law, it remains integrable, and the dynamics are tightly constrained by a complete set of conserved quantities. These are fixed by the initial conditions and differ substantially between the Néel and Vacuum states. Consequently, the time-evolved stationary state inherits nontrivial signatures of the initial state. As already mentioned, the convergence behavior differs between the two preparations, and this may underlie the observed discrepancies in the subleading scaling corrections.

\begin{figure}[t!]
\includegraphics[width=0.5\textwidth]{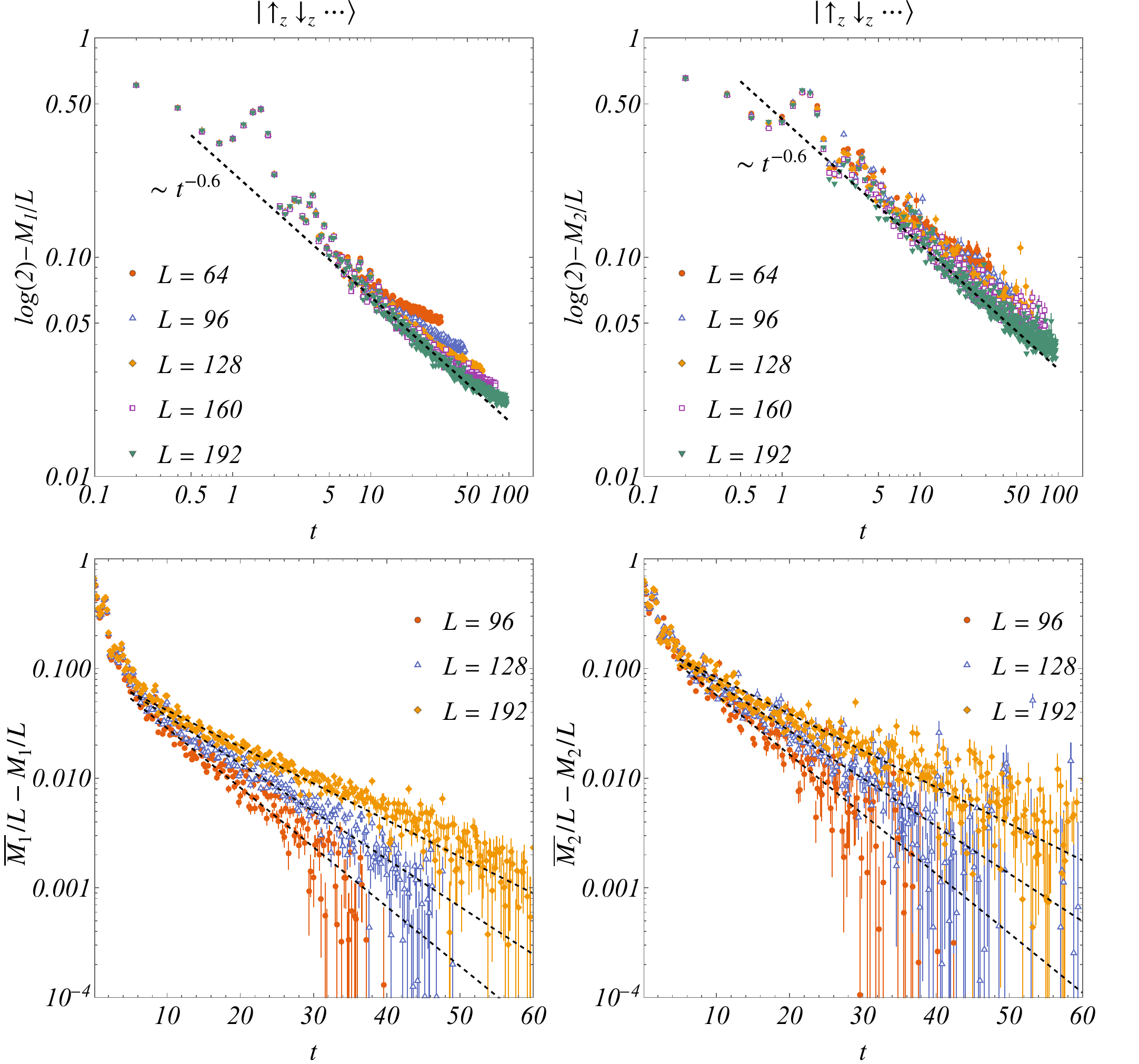}
\caption{\label{fig:ising_plain_neel_time} Time evolution of the stabilizer Rényi entropy density approaching its thermodynamic limit after quenching the N\'eel state with the Ising Hamiltonian at $h=0.5$. \textbf{{Top panels}}. The log-log plot highlights the thermodynamic scaling regime for times $t < t^*(L)$, where finite-size effects are negligible. The dashed line indicates the algebraic decay observed in this regime, as discussed in the main text.
\textbf{{Bottom panels}}. Here, the log-linear plot highlights the exponential decay of the finite-size system towards its best-fit stationary values, given by $\overline{M_{\alpha}(L)} \sim a_{\alpha}L - b_{\alpha} \log L - c_{\alpha}$. The dashed lines represent size-dependent exponential fits, shown as guides to the eye.
}
\end{figure}

For any finite system size $L$, we have seen that the stationary values of the SREs display logarithmic corrections to the leading extensive behavior. This naturally prompts the question: how do the SREs dynamically approach this asymptotic form at large times? Consistent with earlier findings from both random circuit dynamics and Gaussian evolutions~\cite{Turkeshi_2024_2,Collura_2025}, the finite-size convergence to stationarity can follow an exponential decay. 

Although we have observed that the dynamics from the Vacuum state tend to approach their finite-size stationary values more rapidly, obtaining sufficiently accurate and converged results in that case would require higher precision and significantly longer simulations fro larger system sizes. For this reason, in the following time-dependent analysis, we only focus on the N\'eel initial state, where the stationary leading extensive behavior can be reliably inferred to follow $\overline{M}_\alpha \sim L \log(2)$.

To examine this, we analyze the difference between the time-dependent SREs and their best-fit stationary values, normalized by the system size, namely $(\overline{M_{\alpha}(L)} - M_{\alpha}(t, L))/L$ (see Fig.~\ref{fig:ising_plain_neel_time}). The stationary values are extracted from fits accounting for extensive ($\sim L$), logarithmic ($\sim \log L$), and decaying corrections ($\sim L^{-1}$). The data show a clear exponential approach to the finite-size stationary values, confirming that the $t \to \infty$ limit is taken before the thermodynamic one.

However, as $L$ increases, the curves progressively bend on a log-linear scale and begin to collapse in the early-time regime. This marks the onset of a different universal behavior associated with the thermodynamic limit, where $1 \ll t \ll L$. This transition in scaling is particularly visible for the case $\alpha = 1$.

\begin{figure}[t!]
\includegraphics[width=0.5\textwidth]{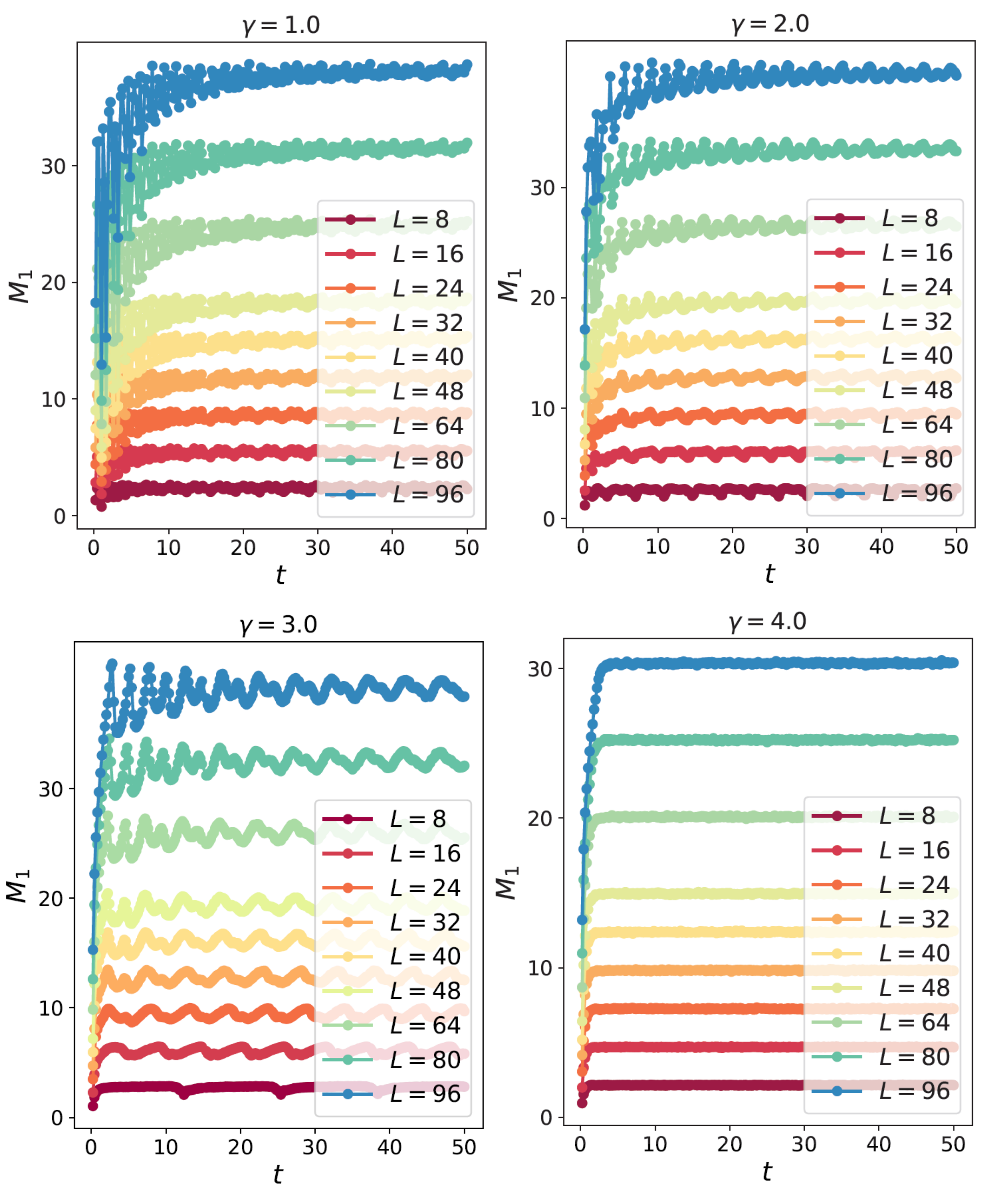}
\caption{\label{fig:IsingNH_time} Time evolution of the SRE $M_1$ under non-unitary free-fermion dynamics for different value of $\gamma$. The system is initialized in a fully polarised state $|\!\uparrow_z \cdots \uparrow_z\rangle$. The entropy $M_1$ exhibits a dependence on $L$ for small and big values of $\gamma$. 
}
\end{figure}

\paragraph{Quantum jump: no-click limit. ---}
In the following setup, we consider the Ising Hamiltonian with no transverse field 
\begin{equation}
    \hat H=J \sum_{j=1}^L \hat\sigma^x_j \hat\sigma^x_{j+1}
\end{equation}
with open boundary conditions, evolving under the competing effect of its own unitary dynamics and the measurement of $\hat\sigma^z_j$, which is indeed proportional to the fermionic local occupation operator $\hat n_j$. 

One can describe the measurement dynamics using the quantum jump formalism, where measurements occur as rare, abrupt events acting stochastically on the quantum state. In this framework, the evolution of the state is governed by a stochastic Schrödinger equation:
\begin{multline} \label{eq:non_hermitian_ham}
d|\psi(\boldsymbol{N_t})\rangle=-i \hat H dt |\psi(\boldsymbol{N_t})\rangle -\frac{\gamma}{2} dt \sum_i (\hat n_i-\langle \hat n_i \rangle_t)|\psi(\boldsymbol{N_t})\rangle\\
+\sum_i \left(\frac{\hat n_i}{\sqrt{\langle \hat n_i \rangle_t}}-1 \right) \delta N^i_t |\psi(\boldsymbol{N_t})\rangle
\end{multline}
where $\boldsymbol{N_t}$ is a Poisson process, and $\delta N^i_t = 0,1$ indicates whether a measurement "click" occurs at site $i$. The expected number of events per time step is given by $\overline{\delta N^i_t} = \gamma dt \langle \hat{n}_i \rangle_t$. Each trajectory generated by this equation remains pure and provides a particular realization of the monitored dynamics, corresponding to an unraveling of the Lindblad evolution.
If no measurements are registered (i.e., $\delta N^i_t = 0$ for all $i$ and $t$), the system evolves deterministically under the effective non-Hermitian Hamiltonian
\begin{equation}\label{eq:non_hermitian_Ising}
\hat H_\text{eff} = J\sum_{i=1}^{L-1} \hat\sigma^x_i\hat\sigma^x_{i+1} - i \frac{\gamma}{2} \sum_i \hat n_i,
\end{equation}
where the imaginary term encodes the influence of the environment. Conversely, if a click occurs, the wavefunction is projected locally onto an up-polarized state.
We focus on the post-selected regime where no clicks occur during the evolution, often referred to as the no-click limit. In this limit, the dynamics are entirely governed by the non-Hermitian Hamiltonian above. This deterministic evolution offers a complementary perspective to the projective measurement framework. The non-Hermitian approach thus captures distinct features of weak monitoring and reveals rich dynamical behavior in the interplay between unitary evolution and measurement back-action.

\begin{figure}[t!]
\includegraphics[width=0.5\textwidth]{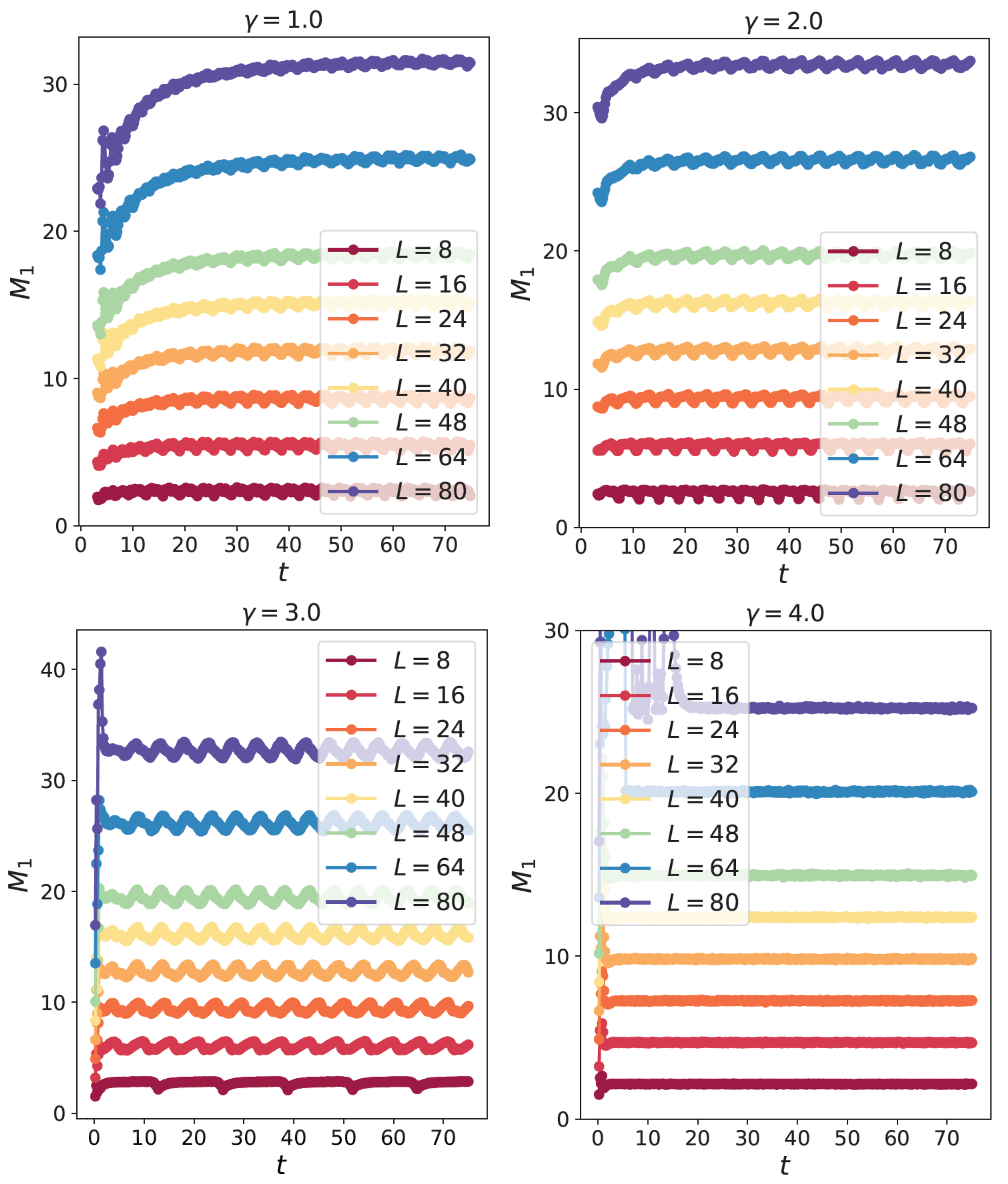}
\caption{\label{fig:IsingNH_time2} Time evolution of the SRE $M_1$ under non-unitary free-fermion dynamics for different value of $\gamma$. The system is initialized in the N\'eel state $|\!\uparrow_z \downarrow_z\uparrow_z \downarrow_z \cdots \rangle$. The entropy $M_1$ exhibits a dependence on $L$ for small and big values of $\gamma$.
}
\end{figure}

We start our analysis from two distinct initial states: the fully polarized state $|\!\uparrow_z \cdots \uparrow_z\rangle$ and the N\'eel state $|\!\uparrow_z \downarrow_z\uparrow_z \downarrow_z \cdots \rangle$, and investigate how varying measurement rates $\gamma$ impact the system’s complexity. For the numerical calculations, we implement the evolution under Eq.~\ref{eq:non_hermitian_Ising} using large-scale matrix-product state simulations, and nonstabilizerness is evaluated using the perfect sampling algorithm of Refs.~\cite{Lami_2023_2,Haug_2023_2}.

Figures \ref{fig:IsingNH_time} and \ref{fig:IsingNH_time2} show the time evolution of the SRE $M_1$, respectively for the initial polarised state (Fig. \ref{fig:IsingNH_time}) and for the N\'eel state (Fig. \ref{fig:IsingNH_time2}). Both figures clearly illustrate that, across different measurement rates $\gamma$, the SRE densities exhibit nontrivial size-dependent dynamics. For low measurement rates, the complexity growth is robust, saturating towards extensive stationary values. However, as $\gamma$ increases, the measurements dominate, suppressing the overall complexity and reducing finite-size effects. These dynamics clearly demonstrate the intricate interplay between unitary complexity generation and measurement-induced complexity suppression.

\begin{figure}[t!]
\includegraphics[width=0.5\textwidth]{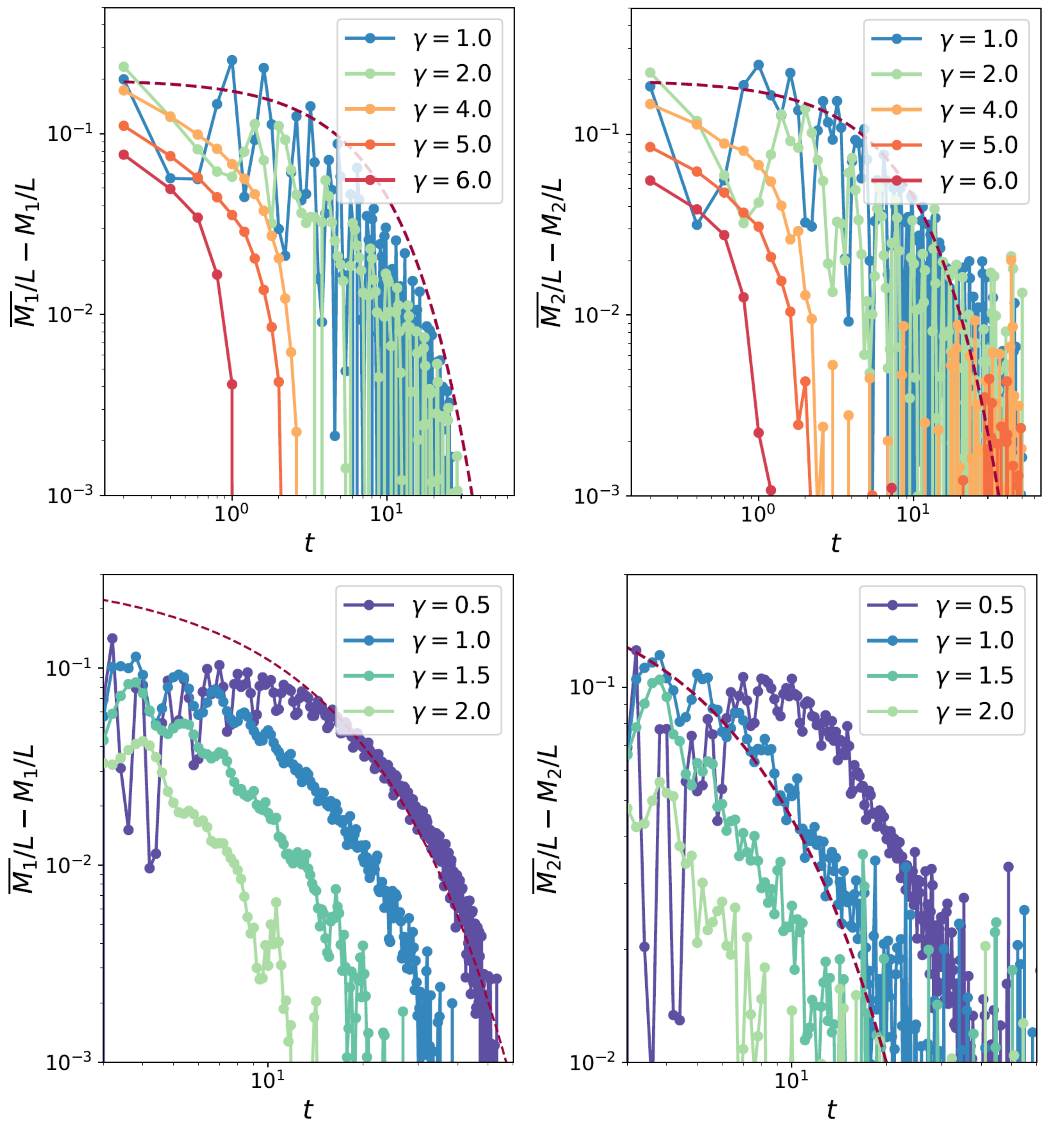}
\caption{\label{fig:nh_plain_power} Log-log plot of the relaxation dynamics of the SREs $M_1$ and $M_2$ after having subtracted the best-fit finite-size stationary values $\overline{M_{\alpha}}(L)$. \textbf{Upper panels} correspond to quenches from the vacuum initial state, while \textbf{lower panels} show the results for quenches from the Néel initial state. In all cases, dashed lines indicate exponential decays
 $e^{-\beta t}$ drawn as guides to the eye. 
}
\end{figure}

\begin{figure}[t!]
\includegraphics[width=0.5\textwidth]{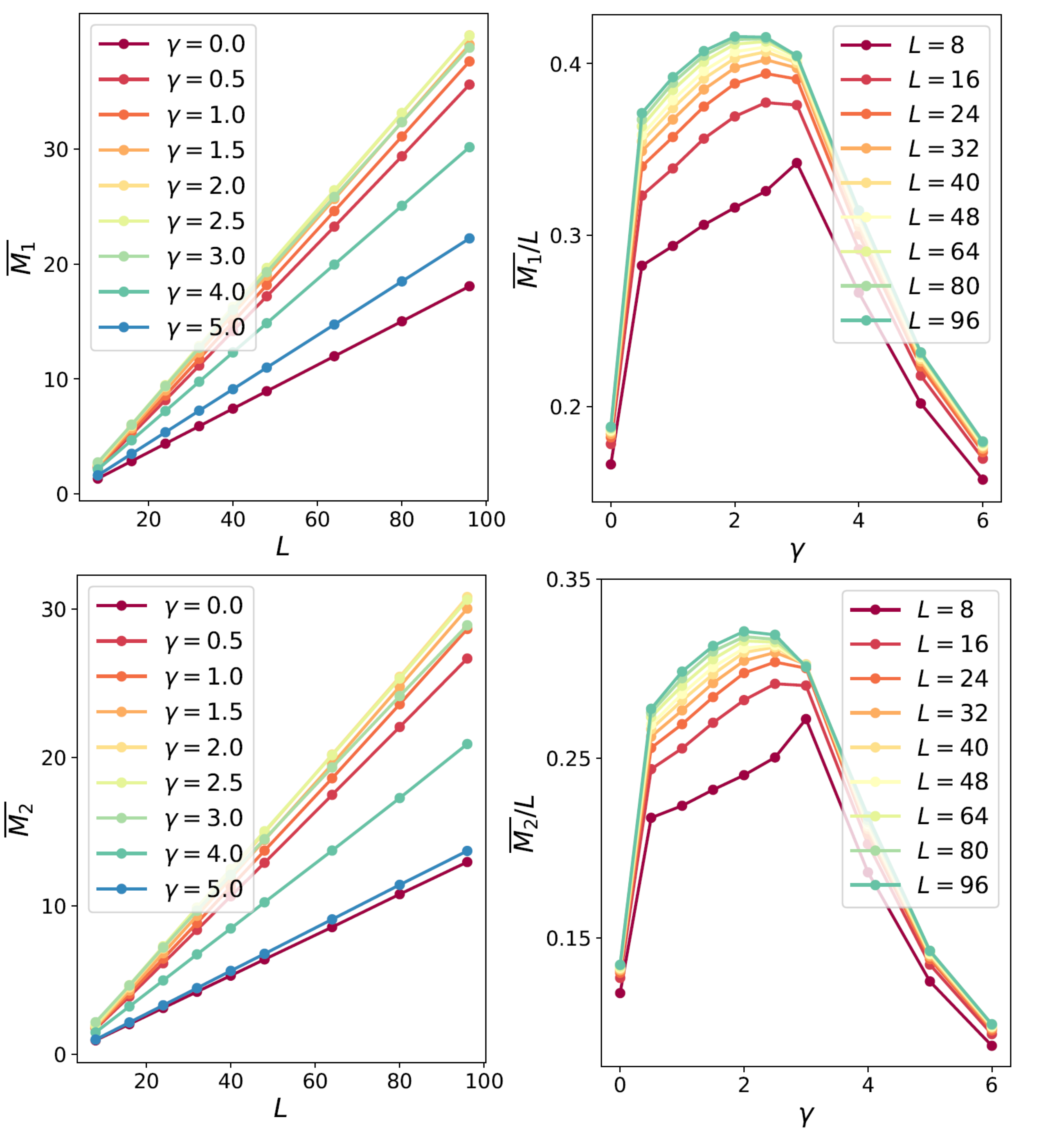}
\caption{\label{fig:nh_vacuum_scaling} Scaling of the stabilizer Rényi entropy densities $\overline{M_1}$ and $\overline{M_2}$ in the stationary regime for various values of $\gamma$. The system is initialized in a fully polarised state $|\!\uparrow_z \cdots \uparrow_z\rangle$. In each panel, the data points represent time-averaged values over a suitable late-time window. \textbf{Left panels}. The entropy densities as functions of the system size $L$, highlighting their extensive scaling behavior. \textbf{Right panels}. The extracted leading coefficients (i.e., slopes from linear fits in $L \in [8,96]$) as functions of $\gamma$, demonstrating that the non-stabilizerness density varies smoothly with the measurement rate without exhibiting abrupt transitions.}
\end{figure}

In Figure \ref{fig:nh_plain_power}, we investigate the relaxation dynamics by considering the deviation from stationary values, $\Delta M_{\alpha}(t,L)=\overline{M_{\alpha}}(L)-M_{\alpha}(t,L)$ in a log-log scale. For all considered values of $\gamma$, an exponentially decay $e^{-\beta t}$, is observed, characterized by different decay exponents dependent on the measurement strength. This behavior highlights that the measurement-induced dynamics not only suppress stationary complexity but also alter relaxation rates.

Figure~\ref{fig:nh_vacuum_scaling} summarizes the stationary regime by displaying the SRE entropies $\overline{M_1}$ and $\overline{M_2}$ as functions of system size $L$ and measurement rate $\gamma$. Here the system is initialized in a fully polarised state $|\!\uparrow_z \cdots \uparrow_z\rangle$.
For small but finite values of $\gamma$, the extensive coefficient of the SRE rises sharply, signaling a rapid enhancement in magic production. This is consistent with the fact that, in the absence of monitoring ($\gamma = 0$), the dynamics are governed by the classical Ising Hamiltonian and no transverse field. Although this Hamiltonian is non-diagonal in the computational basis, it generates trivial, non-entangling dynamics from product states and produces limited magic. Turning on $\gamma$ introduces an effective imaginary transverse field via the no-click evolution, rendering the dynamics nontrivial and enabling efficient magic generation. However, as $\gamma$ increases further, the effect of strong measurements in the Pauli basis dominates, progressively collapsing the wavefunction and thereby suppressing the generation of non-stabilizerness (right panels of Fig.~\ref{fig:nh_vacuum_scaling}). This results in a reduced magic density and a weakened dependence on system size in the stationary regime. Notably, this suppression is continuous, with no abrupt transitions in the leading extensive terms; instead, we observe a smooth crossover. Moreover, in the left panels of Fig.~\ref{fig:nh_vacuum_scaling}, we show the raw time-averaged data 
for different values of $\gamma$, demonstrating the extensive behaviour of non-stabilizerness.

In contrast, as highlighted by the finite-size scaling analysis for the previous cases, the subleading logarithmic corrections exhibit sharp transitions at critical measurement rates. This behavior signals subtle yet robust measurement-induced complexity transitions. To explore this further, we turn to a detailed analysis of the subleading scaling behavior under non-Hermitian evolution in the no-click limit, since we expect that, also in this monitored setup, the stationary SREs exhibit logarithmic corrections of the form $\overline{M}_{\alpha}= a_{\alpha} L -b_{\alpha} \log L -c_{\alpha}$. To probe this subleading structure, we again analyze the finite-size difference $\overline{M}_{\alpha}(2L)-2 \overline{M}_{\alpha}(L)$.

\begin{figure}[t!]
\includegraphics[width=0.5\textwidth]{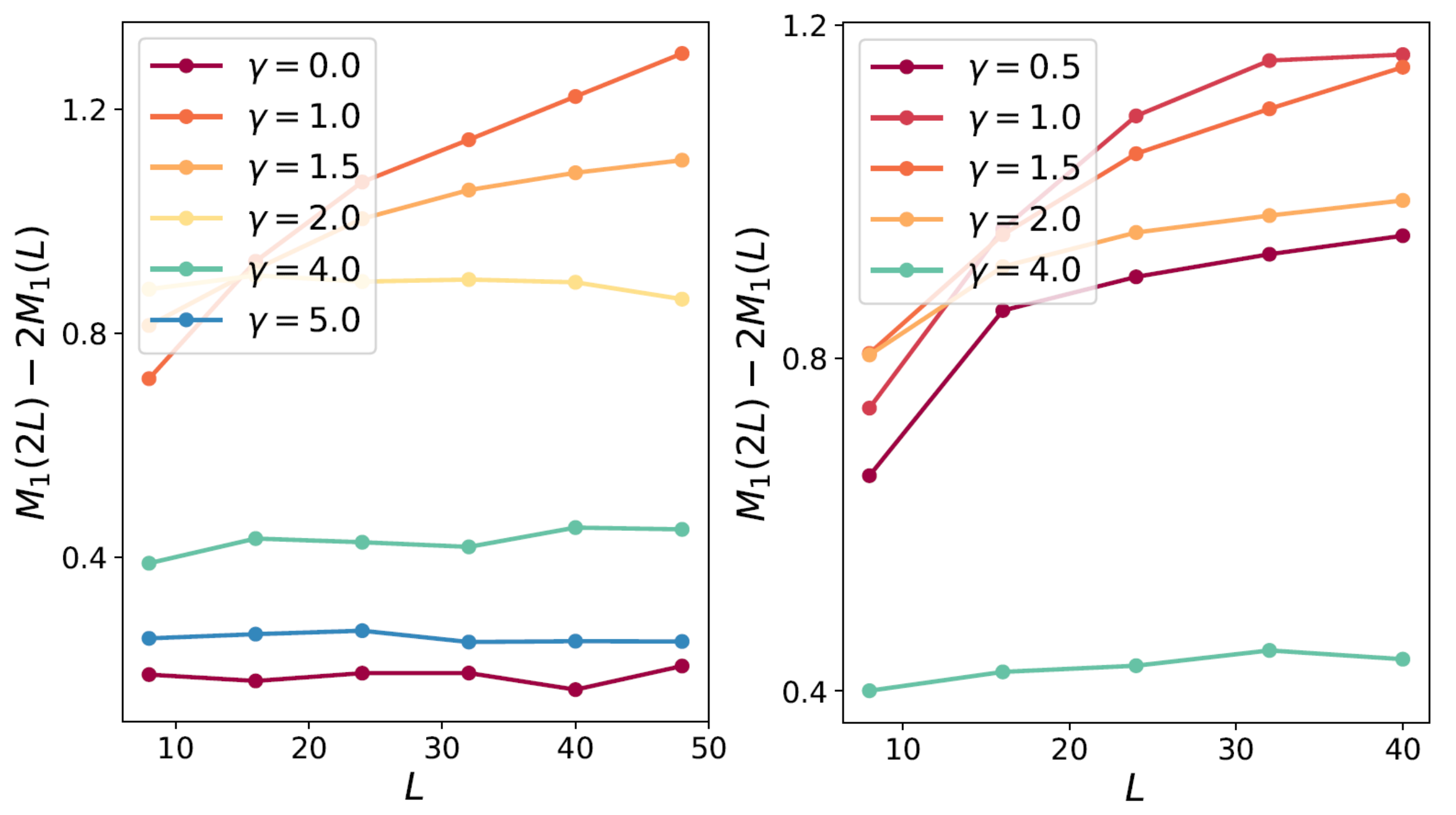}
\caption{
\label{fig:ising_nh_log}  
Finite-size analysis of the corrections to stationary stabilizer Rényi entropies under non-Hermitian Hamiltonian evolution (no-click limit), contrasting vacuum (left panel), and Néel (right panel) initial states. Both panels show the finite-size difference $\overline{M}_1(2L)-2\overline{M}_1(L)$ as a function of system size $L$, for different measurement rates $\gamma$. At small $\gamma$, data show a clear logarithmic growth with $L$, indicating robust logarithmic corrections to complexity.  For $\gamma > 3$ the finite-size difference becomes independent of $L$, signaling the disappearance of logarithmic corrections and thus a simplification of the complexity structure.
}
\end{figure}

Figure~\ref{fig:ising_nh_log} provides a detailed analysis of the subleading logarithmic corrections to the SRE for different measurement rate $\gamma$, comparing the vacuum (left panels) and Néel (right panels) initial states.   
Both panels show the finite-size differences plotted against $L$, clearly showing a $\log L$ dependence, for small measurement rates. This is strong evidence of persistent logarithmic corrections to the stationary non-stabilizerness in the weakly monitored regime.
Crucially, as the measurement rate $\gamma$ increases, the slope of these curves systematically decreases, signaling a suppression of logarithmic corrections.  
For both vacuum and Néel initial states, the data clearly show that the subleading logarithmic term vanishes sharply beyond a threshold measurement rate $\gamma>3$, indicating an abrupt measurement-induced transition in the subleading scaling structure of complexity.

\section{Conclusions and Outlooks}\label{sec:conclusion}
We have investigated the dynamics of SREs as measures of quantum complexity in monitored Gaussian fermionic systems. Our focus is on two paradigmatic models: free hopping fermions and the TFIM, considering both unitary and monitored evolutions.

For free hopping fermions, when examining the full system, the SREs show negative logarithmic corrections to the leading extensive scaling, revealing a universal subleading structure. Introducing measurements triggers a nontrivial measurement-induced transition, characterized by a change in the logarithmic corrections of the SREs. This transition signals a fundamental change in the scaling of complexity, demonstrating that monitoring affects nonstabilizerness in ways that go beyond entanglement transitions alone.

In the TFIM, starting from either Néel or Vacuum initial states, the unitary quench dynamics cause extensive growth of SREs that saturate near the Haar-random limit. Pronounced logarithmic corrections appear, which depend on both the Rényi index and the choice of initial state. The integrability of the model and its conserved quantities imprint subtle initial-state dependence on these corrections.

When stochastic projective measurements in the $\hat{\sigma}_z$ basis are introduced at a rate $\gamma$, and the system evolves under an effective non-Hermitian Hamiltonian capturing the no-click limit of quantum jump trajectories, the dynamics become markedly richer. The interplay between unitary Ising couplings and measurement effects—encoded in the non-Hermitian evolution—shapes the growth and saturation of the stabilizer Rényi entropies, giving rise to measurement-induced complexity transitions. Our numerical simulations reveal a pronounced system-size dependence of the stationary SREs across varying measurement rates, uncovering distinct dynamical phases characterized by different scaling behaviors of the subleading terms of magic.

Finally, the time-dependent analysis in both models reveals an exponential approach to finite-size stationary values, followed by a crossover to an algebraic thermodynamic scaling behavior when $1\ll t \ll L$.

Together, our results demonstrate that monitored Gaussian fermionic systems exhibit rich dynamical and stationary complexity phenomena that extend beyond entanglement properties alone. SREs prove to be powerful tools for probing these complexity transitions, which sensitively depend on integrability, measurement strength, and initial conditions.

Looking ahead, developing analytical frameworks to understand these logarithmic corrections and measurement-induced complexity transitions will be crucial. Extending these investigations to interacting and non-Gaussian systems, and exploring their implications for quantum error correction and simulation, represent promising future research directions.

\subsection*{Acknowledgments}
The authors acknowledge valuable discussions with Angelo Russomanno, Jacopo De Nardis, and M. Dalmonte.
E.T. and X.T. acknowledge collaborations and discussions with P. Sierant and P.S. Tarabunga on related subjects.
X.T. acknowledges DFG Collaborative Research Center (CRC) 183 Project No. 277101999 - project B01 and DFG under Germany's Excellence Strategy – Cluster of Excellence Matter and Light for Quantum Computing (ML4Q) EXC 2004/1 – 390534769. 
M. C. acknowledges support from the PNRR MUR project PE0000023-NQSTI, and the PRIN 2022 (2022R35ZBF) - PE2 - ``ManyQLowD''.
E.\,T. acknowledges support from  ERC under grant agreement n.101053159 (RAVE), and
CINECA (Consorzio Interuniversitario per il Calcolo Automatico) award, under the ISCRA 
initiative and Leonardo early access program, for the availability of high-performance computing resources and support.

\subsection*{Author Contributions}
M.C. supervised the research.
E.T. and M.C. conceived the initial idea and carried out the numerical simulations and data analysis. 
All authors contributed to the development of the project, discussed the results, and contributed to writing and revising the manuscript.

\appendix

\section{Stationary Magic within the Generalized Gibbs Ensemble}
\label{sec:appendix_GGE}
A well-known result in the study of quench dynamics of isolated many-body integrable systems is that, at late times, local observables can (formally) be computed as if the system were described by a Generalized Gibbs Ensemble (GGE) (see the review Ref.~\cite{Essler_2016} and references therein).

In the case of a non-interacting fermionic theory, the GGE takes a particularly simple form~\cite{Rigol_2008}
\begin{equation}
\hat{\rho}_{\mathrm{GGE}} = \frac{1}{Z} \exp\left( \sum_k \lambda_k \hat{n}_k \right),
\end{equation}
with $Z = \Tr\left[ \exp\left( \sum_k \lambda_k \hat{n}_k \right) \right]$,
where $\hat{n}_k$ are the mode occupation number operators corresponding to the single-particle eigenmodes that diagonalize the post-quench Hamiltonian, and $\lambda_k$ are the associated Lagrange multipliers fixed by the initial conditions.

This description implies that, for any local observable $\hat{O}_\ell$ supported on a subsystem of size $\ell$, the long-time limit (when it exists) coincides with its time average and can be computed using the GGE. In other words,
\begin{equation}
\lim_{t \to \infty} \langle \hat{O}_\ell(t) \rangle = \Tr(\hat{O}_\ell \, \hat{\rho}_{\mathrm{GGE}}).
\end{equation}
Eventually, when considering the thermodynamic limit—i.e., when the system size tends to infinity—one may also take the subsystem size $\ell$ to scale accordingly. In this extended setting, the above relation is expected to hold in a thermodynamic sense, albeit with some caveats. Specifically, in the previous equation, the limit $L \to \infty$ was taken first, meaning that all quantities were already evaluated in the thermodynamic regime. However, sending $\ell \to \infty$ afterward, while keeping $\ell \ll L$, is not generally equivalent to taking $\ell = L$ from the outset and then letting $L \to \infty$. These two procedures may lead to qualitatively different behaviors. Thus, the setting we consider here explores a different regime compared to that discussed in the main text: we now focus on large but finite subsystems embedded in the thermodynamic bulk, rather than on the global properties of the entire system.

\paragraph{Hopping Fermions.---} 
In the absence of measurements, the plain unitary dynamics of hopping fermions initialized in a generalized Néel state with fixed density $n$ leads, at late times, to a GGE. For any subsystem of size $\ell$, the reduced density matrix approaches a form described by the following restricted Majorana covariance matrix
\begin{equation}
\mathbb{\Gamma}_{\ell} = \Id_{\ell} \otimes 
\begin{pmatrix}
0 & 1 - 2n \\
2n - 1 & 0
\end{pmatrix},
\end{equation}
where $\Id_{\ell}$ is the identity matrix on the $\ell$-site subsystem.

From this structure, it is straightforward to compute the probability associated with a given Majorana string operator $\hat{\gamma}^{\pmb{x}}$. This probability depends only on the number $m_{\pmb{x}}$ of contiguous pairs of Majorana operators appearing in the string, for which the determinant of the corresponding submatrix is non-vanishing. All other configurations yield zero contribution. For the non-vanishing cases, the probability takes the form:
\begin{equation}
\pi_{\ell}(\pmb{x}) = \frac{(1 - 2n)^{2 m_{\pmb{x}}}}{\left[1 + (1 - 2n)^2\right]^{\ell}}.
\end{equation}

The stabilizer Rényi entropies can then be computed directly from the definition, by summing over all such valid contributions. Specifically, one obtains:
\begin{align}\label{eq:gge_magic}
M_{\alpha}(\ell) &= \frac{1}{1 - \alpha} \log \left[ \sum_{m=0}^{\ell} \binom{\ell}{m} \frac{(1 - 2n)^{2 m \alpha}}{\left[1 + (1 - 2n)^2\right]^{\ell \alpha}} \right] - \ell \log 2 \nonumber \\
&= \ell \left[ \frac{1}{1 - \alpha} \log \left( \frac{1 + (1 - 2n)^{2\alpha}}{\left[1 + (1 - 2n)^2\right]^{\alpha}} \right) - \log 2 \right].
\end{align}

\begin{figure}[t!]
\includegraphics[width=0.5\textwidth]{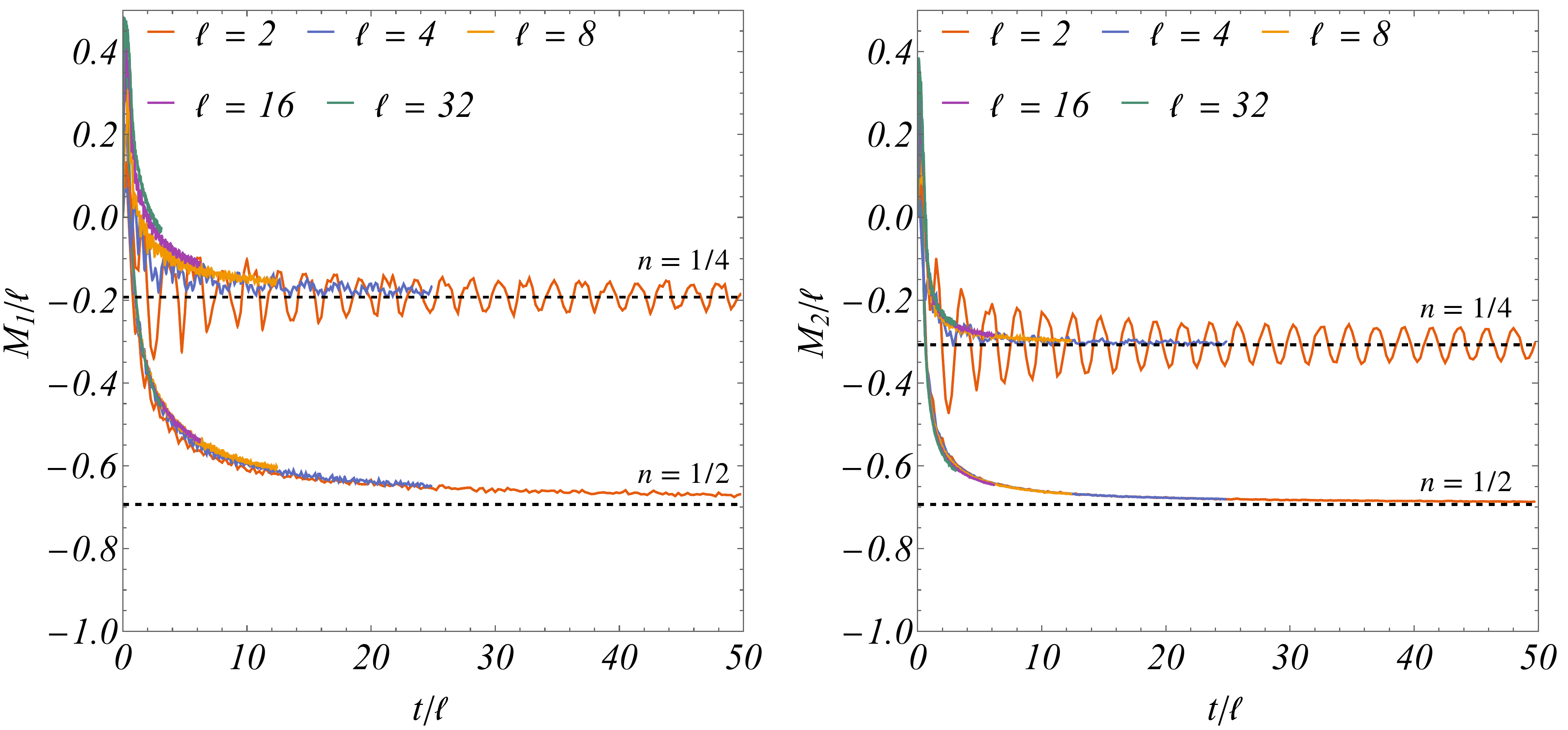}
\caption{\label{fig:XX_neel_gge} Time evolution of the {\it subsystem} stabilizer Rényi entropy densities $M_1/\ell$ (\textbf{left panel}) and $M_2/L$ (\textbf{right panel}) under unitary free-fermion dynamics. The full system of size $L = 256$ is initialized in the N\'eel state at filling $n=\{1/2,1/4\}$. As a function of the rescaled time $t/\ell$ all curves tend to a $\ell$-independent evolution which approaches the infinite-time stationary value (black dashed lines) computed in the GGE. The decay is algebraic and compatible with $t^{-1}$.
}
\end{figure}

In Fig.~\ref{fig:XX_neel_gge}, we present the time evolution of the SREs for subsystems embedded within a larger system of size $L = 256$. We focus on the dynamical regime defined by $\ell \ll t \ll L$, where a Generalized Gibbs Ensemble (GGE) description is expected to hold locally within the subsystem.

After a very short transient—of the order of the subsystem size—during which the subsystem non-stabiliserness increases following a behavior similar to that of the full system (not visible in the figures), the long-time dynamics reveal a qualitatively different trend. In this scaling limit, the SREs of the subsystem decrease toward the GGE value predicted by Eq.~(\ref{eq:gge_magic}), which notably can take negative values. Furthermore, the relaxation toward this stationary value appears to follow an algebraic decay, approximately as $\sim (t/\ell)^{-1}$.

\bibliography{biblio_measurements}

\end{document}